\documentclass{pasj00}

\begin{document}
\SetRunningHead{Hayasaki et al.}{Binary Black Hole Accretion Flows in Merged Galactic Nuclei}
\Received{2006/9/6}
\Accepted{2006/12/15}

\title{Binary Black Hole Accretion Flows in Merged Galactic Nuclei}

\author{Kimitake \textsc{Hayasaki}}

\altaffiltext{}{Yukawa Institute for Theoretical Physics,\\ 
Oiwake-cho, Kitashirakawa, Sakyo-ku, Kyoto 606-8502 }
\email{kimitake@yukawa.kyoto-u.ac.jp}
\author{Shin \textsc{Mineshige}}
\affil{Yukawa Institute for Theoretical Physics,\\ 
Oiwake-cho, Kitashirakawa, Sakyo-ku, Kyoto 606-8502 }

\and 

\author{Hiroshi \textsc{Sudou}}
\affil{Faculty of Engineering, Gifu University, Gifu 501-1193}

%


\maketitle

\begin{abstract}

We study the accretion flows from the circumbinary disks onto 
the supermassive binary black holes 
in a subparsec scale of the galactic center
, using a smoothed particles hydrodynamics (SPH)
code.
Simulation models are presented 
in four cases of a circular binary
with equal and unequal masses, 
and of an eccentric binary with equal and unequal masses.
We find that the circumblack-hole disks are formed 
around each black holes regardless of simulation parameters.
There are two-step mechanisms 
to cause an accretion flow from the circumbinary disk
onto supermassive binary black holes: 
First, 
the tidally induced elongation of the circumbinary disk
triggers mass inflow towards
two closest points on the circumbinary disk from the black holes.
Then, the gas is increasingly accumulated 
on these two points owing to
the gravitational attraction of black holes.
Second, 
when the gas can pass
across the maximum loci of the effective binary potential,
it starts to overflow 
via their two points
and freely infalls to each 
black hole.
In circular binaries,
the gas continues 
to be supplied from the circumbinary disk 
(i.e. the gap between the circumbinary disk 
and the binary black hole is always closed.) 
In eccentric binaries,
The mass supply undergoes
the periodic on/off transitions during one orbital period
because of the variation of periodic potential.
The gap starts to close after the apastron and
to open again after the next periastron passage.
Due to this gap closing/opening cycles,
the mass-capture rates 
are eventually strongly phase dependent. 
This could provide observable diagnosis
for the presence of supermassive binary black holes 
in merged galactic nuclei.

\end{abstract}


\section{Introduction}
\label{sec:intro}
{

There is growing evidence that
most galaxies have the
supermassive black holes at their centers
(\cite{KR}; see also \cite{rees} for a classical review).
The remarkable evidence for 
supermassive black holes
is provided by the existence of a 
gas disk with Keplerian rotation
on a subparsec scale found by the radio observations
\citep{miyoshi} and by the asymmetric iron line
profile discovered by the X-ray observations \citep{tanaka}.
In the vicinity of a supermassive black hole, 
the gas in the disk is heated up and 
produces very high luminosity 
by efficiently transforming their gravitational energy 
into radiation (e.g., \cite{lynden}).
Thus, such rotating gas disks with accretion flow 
are considered as the energy sources of active galactic nuclei (AGNs) 
and quasars.

Recently, it has been widely accepted that 
the supermassive black holes play an important role
not only in the activities
of AGNs and quasars but also in
the formation and global evolution of galaxies 
(\cite{SR}; \cite{KH}; \cite{matteo}; \cite{merritt1}}, 
and references therein).
The discovery of the tight correlation 
between black hole mass and the velocity dispersion 
of the bulge component of galaxies \citep{ferra, geb}
supports the scenario that the black holes have grown up in mass 
through the 
hierarchical galaxy mergers, just as galaxies themselves did.
If this scenario is correct,
the supermassive binary black holes (BBHs) will be inevitably
formed during the course of galaxy mergers \citep{milos1}.

There are actually 
a number of observational indications that some galaxies harbour
supermassive BBHs at their centers. 
Main results are listed as follows
(see also \authorcite{komo1} \yearcite{komo1,komo2});
\begin{itemize}
\item Periodic optical and radio outbursts (e.g., OJ287)
\citep{sill,LV,valtao,valtonen}.
\item Wiggled patterns of the radio jet 
indicating precessional motions 
on a parsec scale \citep{YI,roos2,britzen,AC,LR}. 
\item X-shaped morphology of radio lobes 
\citep{merritt2}. 
\item Double peaked broad emission lines in AGNs
\citep{gaskell,Ho}.
\item Double  compact cores 
with the flat radio spectrum \citep{maness,rodriguez}. 
\item Orbital motion of the compact core \citep{sudou}. 
\end{itemize}

It is believed that the supermassive BBHs evolve mainly
via three stages \citep{begel,yu}.
Firstly, each of black holes sinks independently 
towards the center of the common gravitational potential
due to the dynamical friction with neighboring stars.
When the separation between two black holes
becomes as short as $1\,\rm{pc}$ or so,
angular momentum loss by the dynamical friction 
slows down due to the loss-cone effect and
a supermassive hard binary is formed.
This is the second stage.
Finally, 
when
the semi-major axis of the binary
decreases to less than $0.01\,\rm{pc}$,
gravitational 
radiation dominates and then a pair of black holes, eventually, 
merge into a single supermassive black hole.

\begin{figure}
  \begin{center}
    \FigureFile(80mm,80mm){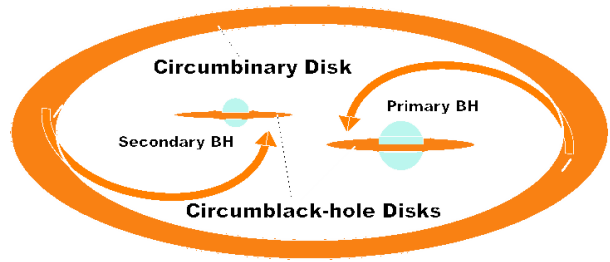}
  \end{center}
  \caption{Schematic diagram of supermassive binary black holes 
           on a parsec/subparsec scale of merged galactic nuclei.
           The supermassive binary black holes 
           are surrounded by a circumbinary disk, from which
           the gas overflows on to the binary, 
           and then the circumblack-hole disks are 
           formed around each black hole (BH).
          }
  \label{fig:system}
\end{figure}

If there is the gas orbiting around the supermassive BBHs
in the second evolutionary stage, 
one will be able to observe a signal arising from
the interaction between the binary and its surrounding gas 
(i.e. a circumbinary disk).
This disk-binary interaction could be also
the predominant candidate to resolve the loss-cone problem 
(\authorcite{armi1} \yearcite{armi1,armi2}; see also \cite{arty1}).
An orbital angular momentum of the supermassive BBHs is transferred to 
the circumbinary disk, by which the gas around the supermassive BBHs 
will be swept away.
In addition, we expect that the mass inflow will 
take place from the circumbinary disk to the supermassive BBHs,
leading to the formation of accretion disks 
around each of black holes (i.e. circumblack-hole disks).
A final configuration could be three-disk systems;
one circumbinary disk and two circumblack-hole disks 
(see Fig.~\ref{fig:system} for a schematic view of supermassive BBHs).

Such three-disk systems have been also discussed
in the context of binary star
formation, where the binary is composed of young stars. 
\citet{arty4} found that the material can infall
on to the central binary through
the gap between the circumbinary disk and the central binary,
and that its accretion rate modulates with the
orbital phase.
Since then, the theory of the young binary star formation has been extensively studied
\citep{arty5,lubow1,bate1,lubow2,GK1,GK2,ochi}.
Some basic processes involved with the disk-binary interaction
have been revealed through these researches.
Despite its significance, however,
it is poorly known how the material 
accretes onto supermassive BBHs
from the circumbinary disk under the realistic situations.
\citet{arty4} only discussed briefly 
the quasi-periodic behavior of 
optical/infrared outbursts
in a blazar OJ287
in terms of the strong phase-dependent accretion onto 
supermassive BBHs (see also \cite{arty1}).

In this paper, therefore,
we elucidate the theory of accretion processes
in supermassive binary black-hole systems by
performing smoothed particle hydrodynamics (SPH) simulations.
Our ultimate goal is to give an observable diagnosis 
for the presence of supermassive BBHs 
in central region of merged galaxies. 
The simulation should ideally take account of all the processes
at work in the three-disk systems, 
including the circumbinary disk evolution 
around the supermassive BBHs,
the mass transfer from the circumbinary disk
to the individual black holes, and the accretion onto 
each of black holes.
Such a simulation, however, 
would require an enormous computational time. 
Therefore, we confine ourselves to simulate only 
the accretion flow from the circumbinary disk
onto the supermassive BBHs. 
Detailed structure and evolution
of the circumblack-hole disks will be reported in a
subsequent paper.

The plan of this paper is as follows:
We first describe our numerical model in Section~\ref{sec:num}.
Then, our numerical results will be reported in Section~\ref{sec:acc}
for the basic processes involved with mass accretion onto the circular,
equal-mass binary, and in Section~\ref{sec:uneq_ecc} for the effects of the
orbital eccentricity and the unequal masses of the binary.
We then discuss the observational implications 
and the related issues in Section~\ref{sec:discussion}.
Section~\ref{sec:conclusions} is devoted to conclusions.

\section{Our Models, Basic Equations, and Numerical Procedures}
\label{sec:num}

\label{subsec:basiceq}
Simulations presented here were performed 
with a three-dimensional (3D) SPH code.
The SPH code is basically the same as that used by
\authorcite{kimi1} (\yearcite{kimi1,kimi2,kimi3}),
and is based on a version originally developed by Benz 
\citep{benz1,benz2}.
The SPH equations with the standard cubic-spline kernel are
integrated using a second-order Runge-Kutta-Fehlberg integrator with
individual time steps for each particle \citep{bate2},
which results in saving an enormous computational time when a large range
of dynamical timescales are involved.
We first give the basic equations and
then describe the implementations to the SPH code.

\subsection{Basic Equations}
\subsubsection{Mass Conservation}
\label{sssec:eqc}

In the SPH calculations, each particle has a density distribution
over a spatial scale of the smoothing length, $h$, around its center.
Hence, the density at the position of particle $i$ ($i=1, 2, \cdots$) 
is given by a weighted summation over the masses of the
particle $i$ itself and its neighboring particles 
(hereafter, neighbors),
\begin{equation}
\rho_{i}= 
\sum_{j}^{N_{\rm{nei}}}m_{j}W(r_{ij},h_{ij}).
\label{eq:1}
\end{equation}
Here $m_j$ is the mass of the particle $j$,
$N_{\rm{nei}}$ is the number of
the neighbors of the particle $i$,
$r_{ij}=|\mbox{\boldmath $r$}_{i}-\mbox{\boldmath $r$}_{j}|$
is the distance between the particles $i$ and $j$,
$h_{ij}=(h_{i}+h_{j})/2$ is the mean smoothing length,
and $W$ is call the kernel.  We adopt
the standard cubic-spline kernel $W$ for 3D; that is,
\begin{eqnarray}
W(r,h) = \frac{1}{\pi{h^{3}}}
\left\{
\begin{array}{ll}
1 - \frac{3}{2}s^{2} +\frac{3}{4}s^{3}
& \rm{if}\hspace{2mm} 0\le{\it s}<1, \\
\frac{1}{4}(2-s)^{3}   & \rm{if}\hspace{2mm} 1\le{\it s}<2, \\
0 & \rm{otherwise},\\
\end{array}
\right.
\label{eq:2}
\end{eqnarray}
where $s\equiv|r_{ij}/h_{ij}|$.
The mass conservation is thus automatically satisfied.

\subsubsection{Momentum Equation}
\label{sssec:eqm}
The momentum equation for the fluid under a gravity
in the inertia frame is written by
\begin{equation}
\frac{d\boldsymbol{v}}{dt} 
= -\frac{\boldsymbol{\nabla}P}{\rho}
-\frac{\boldsymbol{F}_{\rm{vis}}}{\rho}
-\boldsymbol{\nabla}\phi,
\end{equation}
where $d/dt$ is the Lagrangian derivative,
$\boldsymbol{v}$ is the velocity field, $P$ is the pressure,
$\boldsymbol{F}_{\rm{vis}}$ is the viscous force,
and $\phi$ is the gravitational potential by the BBHs.
Self-gravity of the gas particles is neglected.
The corresponding SPH momentum equation for the $i$-th particle
in the potential of a pair of black holes is then
\begin{eqnarray}
\frac{d\boldsymbol{v}_{i}}{dt}
&=&
-\sum_{j}^{N_{\rm{nei}}}m_{j}
\left(
\frac{P_{i}}{\rho_{i}^{2}}+
\frac{P_{j}}{\rho_{j}^{2}}+
\Pi_{ij}
\right)
\nabla_{i}W(r_{ij},h_{ij}) \nonumber \\
&-&\frac{GM_{\rm p}(\boldsymbol{r}_{i}-\boldsymbol{R}_{\rm p})}
    {|\boldsymbol{r}_{i}-\boldsymbol{R}_{\rm p}|^{3}}
 - \frac{GM_{\rm s}(\boldsymbol{r}_{i}-\boldsymbol{R}_{\rm s})}
    {|\boldsymbol{r}_{i}-\boldsymbol{R}_{\rm s}|^{3}}
\label{eq:4}
\end{eqnarray}
where $v_{i}$ is the velocity of the $i$-th particle,
$G$ is the gravitational constant,
$M_{\rm p}$ and $M_{\rm s}$ are the masses of the primary
and secondary black holes, respectively,
$\boldsymbol{R}_{\rm p}$ and $\boldsymbol{R}_{\rm s}$
are the position vectors of the primary and secondary black holes,
respectively,
 and $\Pi_{ij}$ is the SPH artificial
viscosity with the following standard form \citep{mona1},
\begin{eqnarray}
\Pi_{ij}=
\left\{
\begin{array}{ll}
(-\alpha_{\rm{SPH}}c_{\rm{s}}\mu_{ij} + \beta_{\rm{SPH}}\mu_{ij}^{2})
/\rho_{ij} 
& \mbox{\boldmath $v$}_{ij}\cdot\mbox{\boldmath $r$}_{ij}\le0 \\
0          
& \mbox{\boldmath $v$}_{ij}\cdot\mbox{\boldmath $r$}_{ij}>0, \\
\end{array}
\right.
\label{eq:5}
\end{eqnarray}
where $\alpha_{\rm{SPH}}$ and $\beta_{\rm{SPH}}$ are the linear and
non-linear artificial viscosity parameters, respectively.
$\rho_{ij}=(\rho_{i}+\rho_{j})/2$, 
$\mbox{\boldmath $v$}_{ij}=\mbox{\boldmath $v$}_{i}
-\mbox{\boldmath $v$}_{j}$ and
$\mu_{ij}=h_{ij}\mbox{\boldmath $v$}_{ij}\cdot\mbox{\boldmath $r$}_{ij}
/(r_{ij}+\eta_{ij})$
with $\eta_{ij}^{2}=0.01h_{ij}^{2}$.
The connection with the disk viscosity will be described
in subsection 2.2.3.

\subsubsection{Equation of State}
\label{subsec:eos}

The pressure term $P$ in the momentum equation
is calculated by the isothermal equation of state;
\begin{equation}
P_{i}=c_{\rm s}^{2}\rho_{i},
\label{eq:3}
\end{equation}
where $c_{\rm s}$ is the isothermal sound speed of the gas.
We do not need to explicitly solve an energy equation.

\subsection{Disk Viscosity}
\label{subsec:visc}

Viscosity is essential in the disk simulations,
while the SPH formalism already contains the artificial viscosity.
There is
an approximate relation connecting the Shakura-Sunyaev
viscosity parameter $\alpha_{\rm{SS}}$ and the SPH artificial
viscosity parameter $\alpha_{\rm{SPH}}$.
The outline of formalism is presented below 
in the same procedure
as that of Section 2.1 of \citet{oka}.

\citet{MWB} found that the SPH viscous force becomes
\begin{equation}
\mbox{\boldmath $F$}_{\rm{vis}} =
\frac{1}{10}\alpha_{\rm{SPH}}c_{\rm{s}}h
[\mbox{\boldmath $\nabla$}^{2}\mbox{\boldmath $v$} +
2\mbox{\boldmath $\nabla$}(\mbox{\boldmath $\nabla$}\cdot
\mbox{\boldmath $v$})]
\label{eq:9}
\end{equation}
in the 3D continuum limit of equation~(\ref{eq:5}),
assuming that the density varies on a length-scale much longer than
that of the velocity.
This implies that the shear viscosity $\nu$
and the bulk viscosity $\nu_{\rm bulk}$ are given by
\begin{equation}
\nu=\frac{1}{10}\alpha_{\rm{SPH}}c_{\rm{s}}h
\label{eq:10}
\end{equation}
and
\begin{equation}
\nu_{\rm{bulk}}=\frac{5}{3}\nu,
\label{eq:11}
\end{equation}
respectively.

According to the $\alpha$ viscosity prescription \citep{SS},
on the other hand, the shear viscosity is written as
\begin{equation}
\nu=\alpha_{\rm{SS}}c_{\rm{s}}H,
\label{eq:12}
\end{equation}
where $H$ is the disk scale-height and $\alpha_{\rm{SS}}$ is
the viscosity parameter.
>From the combinations of equations~(\ref{eq:10}) and (\ref{eq:12}),
we find the relation connecting 
$\alpha_{\rm{SPH}}$ and $\alpha_{\rm{SS}}$ as
\begin{equation}
  \alpha_{\rm{SS}} = \frac{1}{10}\alpha_{\rm{SPH}}\frac{h}{H},
  \label{eq:13}
\end{equation}
as long as 
$\mbox{\boldmath $\nabla$}\cdot\mbox{\boldmath $v$}=0$ holds.
However, we usually find 
$\mbox{\boldmath $\nabla$}\cdot\mbox{\boldmath $v$}\neq0$
in general flows.
Moreover, the viscosity is artificially tuned off for divergent flows
in our model (see equation~\ref{eq:5}).
Therefore, equation~(\ref{eq:13}) should be taken as a rough approximation
to relate $\alpha_{\rm{SPH}}$ to $\alpha_{\rm{SS}}$.
We adopt a constant value of
$\alpha_{\rm{SS}}=0.1$ throughout the simulations.
Hence,
$\alpha_{\rm{SPH}} =10 \alpha_{\rm{SS}} H/h$ is
variable in space and time, while $\beta_{\rm{SPH}}=0$ everywhere.

\subsection{Initial Settings}
\label{subsec:ic}

We put a pair of black holes on the $x$-$y$ plane with the
semi-major axis of the binary orbit being along the $x$-axis initially
and the center of mass being at the origin.
In the case of an eccentric binary, we set
a pair of black holes initially with the minimum separation
(i.e., at the periastron).
That is, black holes are initially at $(x,y)=(a(1-e),0)$
and at $(x,y)=(-a(1-e),0)$, where $a$ is the semi-major axis
and $e$ is the eccentricity.  In addition, each black hole
is given an appropriate initial rotation velocity so as to
orbit around the origin with given $a$ and $e$.
In the present study, we fix $a=0.1 \rm{pc}$.
As for the eccentricity, we calculate two cases; $e=0$, and 0.5.
The unit of time is $P_{\rm{orb}}\simeq296\,\rm{yr}$,
unless noted otherwise.

We assume that the black holes are both Schwarzschild black holes.
The masses of the primary and the secondary black holes
are $M_{\rm{p}}$ and $M_{\rm{s}}$, respectively.
The binary has the mass ratio $q=M_{\rm{s}}/M_{\rm{p}}$ and
the total mass
$M_{\rm{\rm tot}}\equiv M_{\rm{\rm p}} + M_{\rm{\rm s}} =10^{8} M_{\odot}$.
The black holes are modeled by sink particles with
the fixed accretion radius of
$r_{\rm{acc}}=0.2a$ which are $\sim8.0\times10^{3}$ times as large as
the Schwarzschild radius.
Numerically, we take away all the particles which enter the region
inside $r_{\rm acc}$.
To ensure that the simulation results
do not depend on the accretion radius,
we also performed a simulation with
$r_{\rm{acc}}=0.1a$, finding no qualitative and quantitative
differences within estimated errors of $\sim13\%$.

The circumbinary disk is initially
set around the common center of mass of the supermassive BBHs,
which is coplanar with the binary orbital plane.
It has a radially random density profile over a width of $0.05 a$
and a vertically isothermal, thin disk density profile.
Its initial mass is $1.0\times10^{-4}M_\odot$.
The disk temperature is assumed to be
$T=5000\rm{K}$ everywhere. Note that this temperature
roughly corresponds to the typical effective temperature of a standard disk
around a single black hole with $10^{8}M_{\solar}$ \citep{kato}.
The disk material is rotating around the origin
with the Keplerian rotation velocity.
The gas particles are added randomly at the radius of
the initial outer edge of the circumbinary disk, 
at a constant rate, $\dot{M}_{\rm{inj}}=1.0\,M_{\odot}\rm{yr}^{-1}$
in all the calculated models.

The inner edge of the circumbinary disk 
depends on the orbital eccentricity.
In the circular binary,
we take the radius of $r=1.85a$ corresponding to the tidal
truncation radius where
the tidal torque of the binary equals
to the viscous torque of the circumbinary disk.
A circumbinary disk around a circular binary is truncated
at this radius \citep{PP}.
The tidal truncation radii for circular binaries with
non-extreme masses are distributed
between $r/a=1.68$ and $r/a=1.78$
(see Table~1 of \cite{arty3}).
In the eccentric binary, on the other hand,
we take the (2,1) corotation radius at $r=2.75 a$
(see section \ref{subsec:overview2} for the detailed explanation).

We set an outer calculation boundary at $r = 6.0a$,
which is sufficiently far from the disk region so that
the outer boundary should not affect the flow dynamics 
in the supermassive BBH systems.
The SPH particles passing outward across the outer boundary
are removed from the simulation.

\subsection{Simulation Implementations}
\label{subsec:si}

In our code, the accretion flow is modeled by
an ensemble of gas particles, each of which has a negligible mass
chosen to be $1.0\times10^{-7}M_{\odot}$ with a variable smoothing length.
We have carried out several simulations with different parameters.
The parameters adopted by the calculated models are summarized in Table~1.

To check if the number of SPH particles used in this study is large enough,
we also performed the same simulation as in model~1
but with about a half as many particles
, finding no appreciable changes.
In order to see the validity of the simulations,
we also checked several simulation values, such as
the ratio of the smoothing length to the disk
scale-height $h/H$, the ratio of the smoothing length to
the disk radius $h/r$, and the relative disk scale-height $H/r$, respectively.
Here the disk scale-height $H$ is defined as the half thickness
at which the density decreases by a factor of $e^{-1/2}$.
We have found $h/r \ll 1$ in the range of $1.68a\le{r}\le4.0a$;
that is, the radial structure of the circumbinary disk is well resolved.
We also found $h/r \sim 0.03$ at the inner edge of the disk,
which ensures the justification of
the size of accretion radius, $r_{\rm{acc}}=0.2a$.
The disk is geometrically thin because of
$H/r<0.01$ over the whole radial region,
as expected from the standard disk theory.
However, the vertical structure of the disk
is not resolved by our SPH simulations, since we find $h/H>1$.
We, hence, focus our discussion on the radial structure
and the detailed explanation of vertical structure is
beyond the scope of the present paper.

\begin{table*}
\begin{center}
\caption{
  Summary of model simulations.
  The first column represents the model numbers. 
  The second column is the run time
  in units of $P_{\rm{orb}}$.
  The third column is
  the number of SPH particles at the end of the run.
  The mass ratio and the eccentricity are given in the fourth column and the fifth
  column,respectively.
  The last column is the initial radius of the inner edge of the circumbinary disk.
}
\label{tbl:models}
\begin{tabular}{@{}lccccc}
\hline
Model       & Run time
            & $N_{\rm{SPH}}$
            & Mass ratio
            & Eccentricity
            & Initial disk-inner edge \\
            & $(P_{\rm{orb}})$
            & (final)
            & $q$
            & $e$
            & $r_{\rm{edge}}/a$ \\
\hline
1           & $40$
            & 64023
            & $1.0$ & $0.0$
            & $1.68$ \\
2           & $60$
            & 92633
            & $1.0$ & $0.5$
            & $2.75$ \\
3           & $40$
            & 73975
            & $0.5$ & $0.0$
            & $1.75$ \\
4           & $60$
            & 64921
            & $0.5$ & $0.5$
            & $2.75$ \\
\hline
\end{tabular}
\end{center}
\end{table*}


\section{Circular Binary with Equal-Mass Black Holes}
\label{sec:acc}

In this section
we consider the accretion onto the circular, equal-mass binary
from the circumbinary disk (model~1)
for understanding its basic characteristics.

\subsection{Overall Evolution}
\label{subsec:overview1}

\begin{figure*}
  \begin{center}
    \FigureFile(80mm,60mm){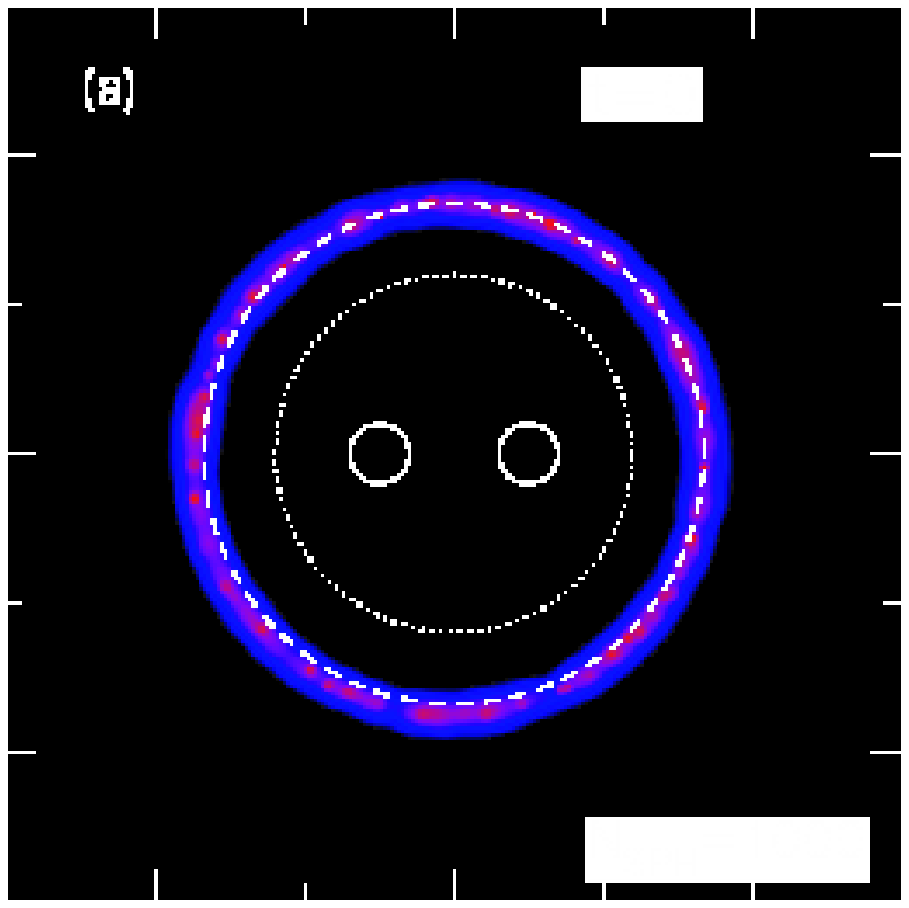}
    \FigureFile(80mm,60mm){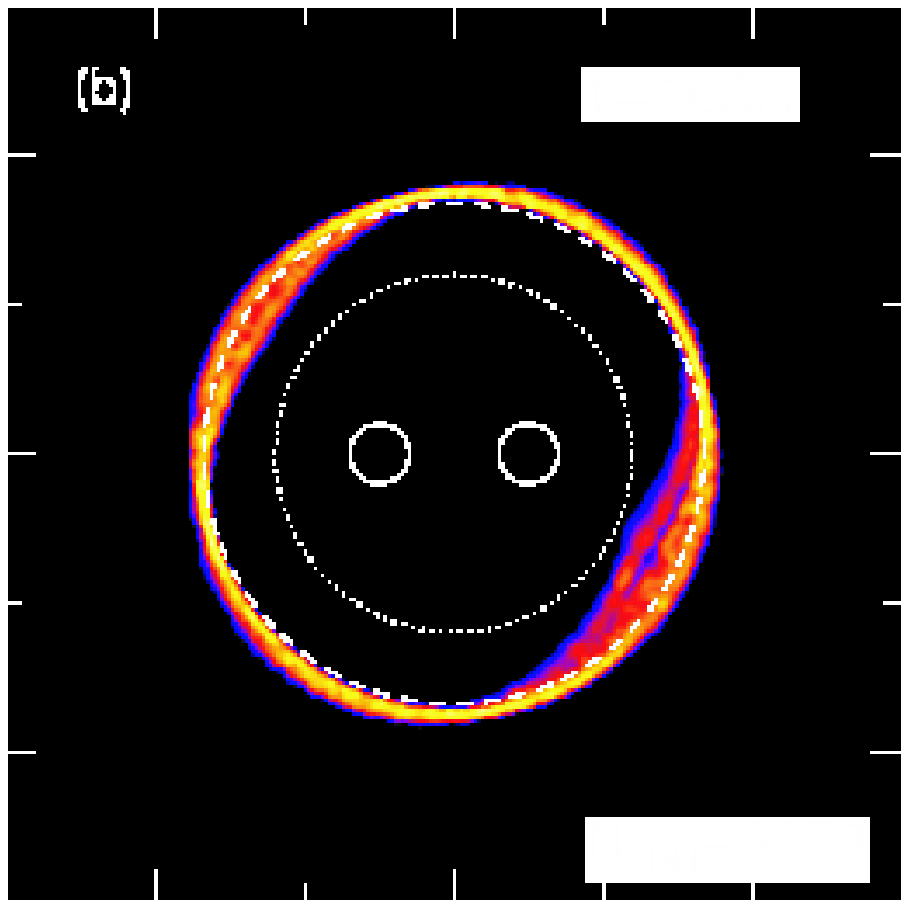}
  \end{center}
  \begin{center}
    \FigureFile(80mm,50mm){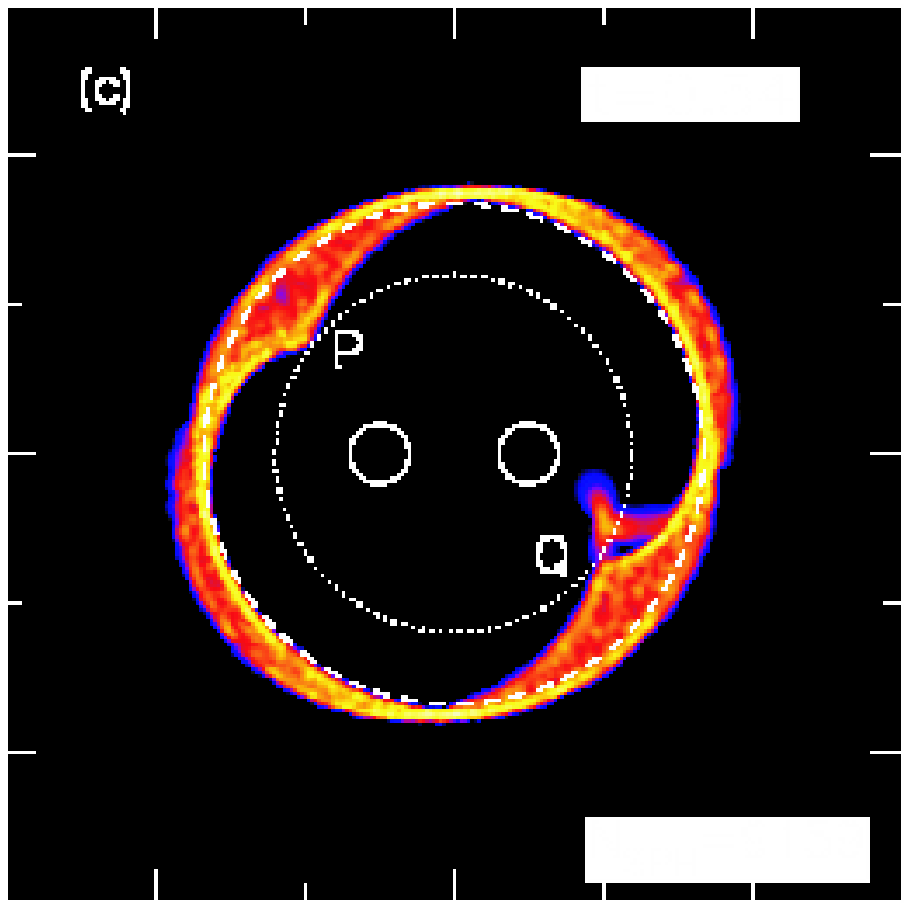}
    \FigureFile(80mm,50mm){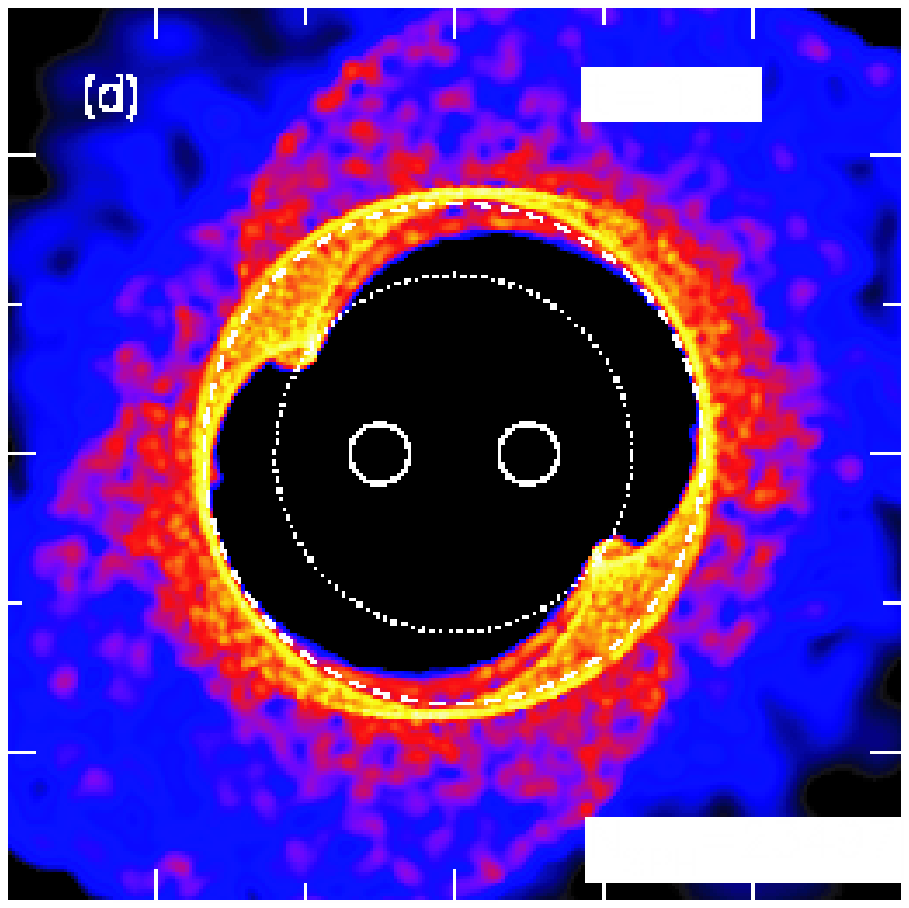}
  \end{center}
  \begin{center}
    \FigureFile(80mm,50mm){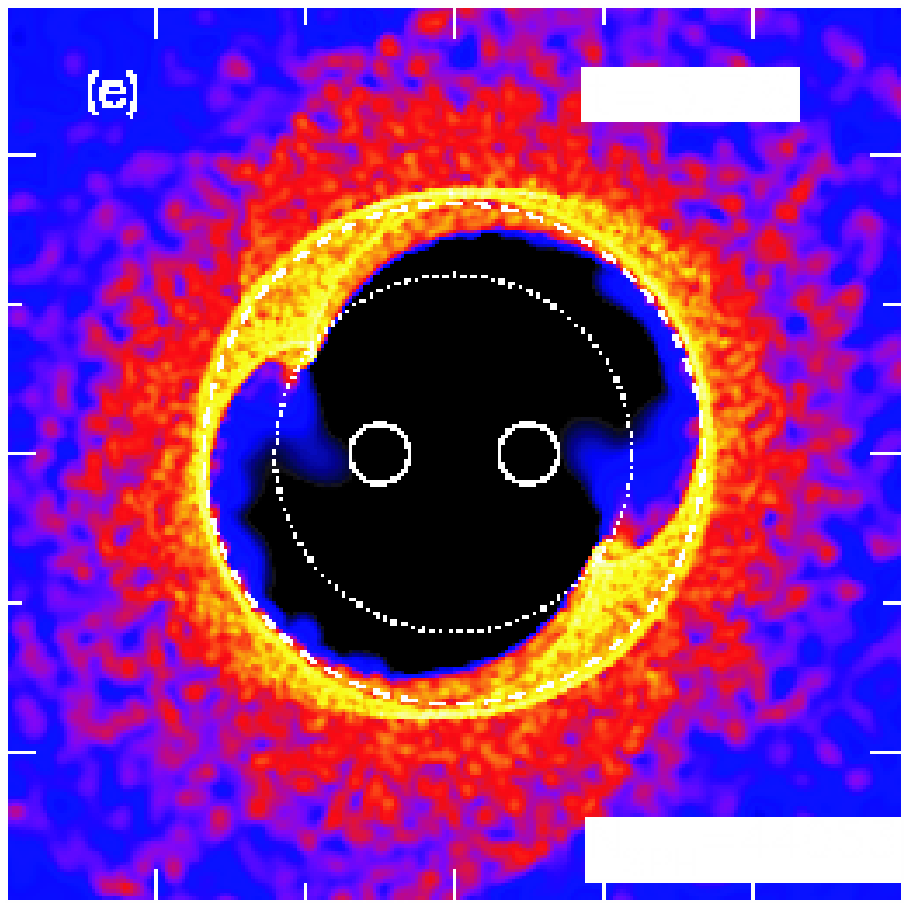}
    \FigureFile(80mm,50mm){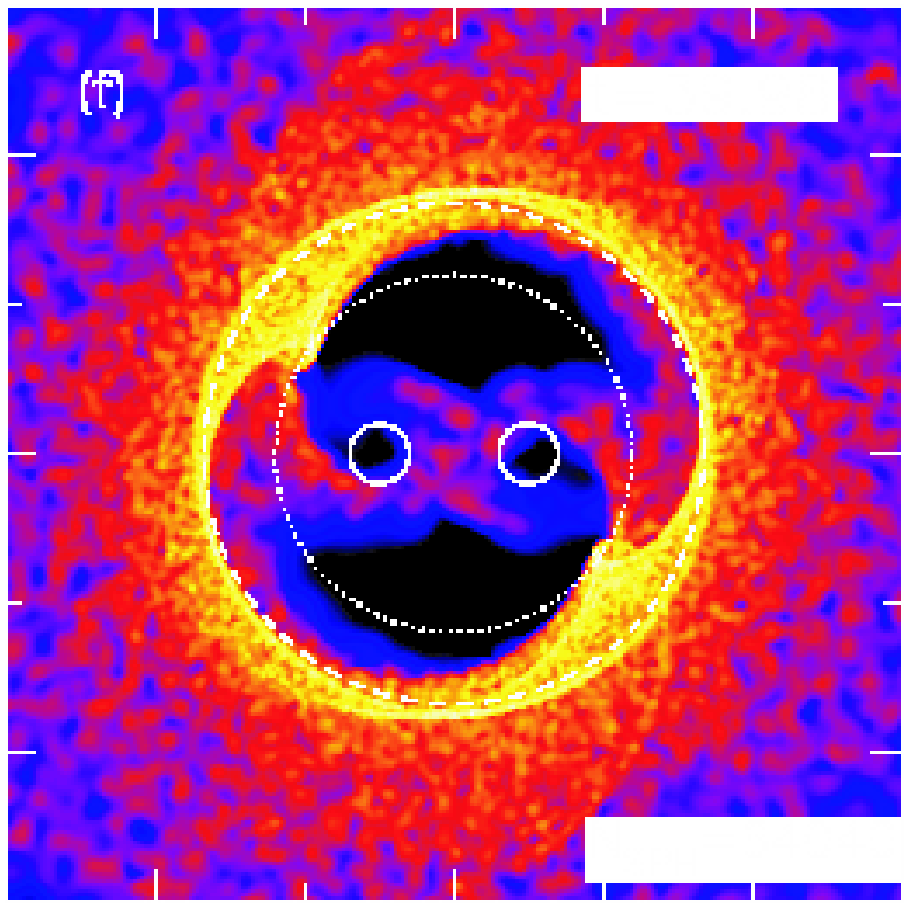}
  \end{center}
\caption{
           Snapshots of accretion flow from the circumbinary disk onto the supermassive BBHs
           in model~1.
           The origin is set on the center of mass of the binary and all panels
           are shown in a binary rotation frame.
           Each panel shows the surface density in a range of three orders
           of magnitude in the logarithmic scale.
           The solid circle denotes the accretion radius of the primary black hole
           and the secondary black hole, respectively.
           The dashed and dotted circle represents the tidal truncation radius
           and the outer Roche-lobe radius of $q=1.0$
           along with $L_{2}$ and $L_{3}$, respectively.
           Annotated in each panel are the time in units of $P_{\rm{orb}}$
           and the number of
           SPH particles $N_{\rm{SPH}}$.
         }
 \label{fig:snap_circ}
\end{figure*}

We first overview the global evolution of the supermassive BBH systems.
Fig.~\ref{fig:snap_circ}
gives snapshots of the accretion flow in 6 evolutionary stages.
These are density contours in the rotation frame co-rotating
with the supermassive BBH.
Both of the black holes and the circumbinary disk
are rotating in the anti-clockwise direction, although
the former are rotating more rapidly than the latter
according to the Kepler's law.
A pair of the solid circles denote the accretion radii
of black holes, which we set at $r_{\rm{acc}}=0.2a$
 from the center of the black holes.
The dotted circle and dashed circle represent
the tidal truncation radius, $r_{\rm{trunc}}=1.68 a$
(see section 2.3), and
a trapping radius for material, $r_{\rm{trap}}/a=1.19841$,
respectively.
The trapping radius is defined as the distance from the
center of mass to the outer Lagrange point, $L_{2}$.
This is also the same as that from the center of mass
to the $L_{3}$ point
in the case of a circular binary with equal masses.
This circle approximately points the loci
of the maximum in the radial profile of the effective potential
(e.g. \cite{kitamura}).
In other words, when the material once flows inside this circle,
it can freely fall
towards either of the black holes.

Let us see the disk evolution, following each panel
in Fig.~\ref{fig:snap_circ}.
Although starting with a circular shape around the center of mass
at $t=0$ [panel (a)],
the disk shape begins to be elongated [see panel (b)
at $t=0.35$].
Such deformation continues to grow as the time goes on,
and eventually the inner edge of the circumbinary disk
touches the circle of the trapping radius
at two points at $t=0.54$ [see points P and Q in panel (c)].
Then, gas starts to flow towards the black holes
across these two points (particularly, via the point Q), 
being gravitationally attracted by the black holes.
But this inflow is only a transient one for $0\le{t}<{1}$.
Subsequently, the  gas is gradually accumulated on the two points 
as shown in panel (d).
After $t=3.75$, the gas starts to continuously overflow via points P and Q.
The gas inflow will eventually form an accretion disk
(i.e., circumblack-hole disks) around each black hole.
The possibility of circumblack-hole disk formation
will be discussed in section~\ref{subsec:diskform}.


\subsection{Why is the Disk Elongated?}
\label{subsec:diskelongation}

Fig.~\ref{fig:snap_circ} clearly demonstrates that
the disk elongation triggers mass inflow towards the black holes.
Then, what is the key physics underlying the disk deformation?
To understand the physics, we
performed the pressure-less particle simulation,
in which we dropped all the terms related to pressure and viscosity
($P_i = \Pi_{ij} = 0$ in equation \ref{eq:4}),
adopting the same parameters as those in model~1.
The resultant evolution is very much similar to those we obtained
in model 1, including the angle of the elongation,
in the initial evolutionary stage for $0<t<1$.
This examination unambiguously proves 
that the key physics
causing the disk elongation should be of kinematics origin; that is,
the potential of the supermassive BBHs
is dominated by the $m=2$ Fourier component of the binary potential.
This $m=2$ Fourier component causes the $m=2$ mode to be excited
on the circumbinary disk, which eventually makes
the circumbinary disk elongated. 

Why is, then, the semi-major axis of the elongated disk misaligned
with the line connecting the primary and the secondary black holes
as seen in panel (b) of Fig.~\ref{fig:snap_circ}?
In the present case, both the binary and the circumbinary disk
rotates with the different angular frequency, where
the angular frequency of the binary is always faster than
that of the circumbinary disk.
The system evolves towards the synchronous rotation
with the resonance between the angular frequency of 
the binary and that of the circumbinary disk, which causes the dissipation on
the circumbinary disk.
Therefore, we interpret that the misalignment is due to the
the resonance friction induced by the difference 
between the binary frequency
and the frequency of the disk-inner edge.

After the disk starts to be elongated,
the disk material is accumulated in the two closest points
on the circumbinary disk by
the gravity force of black holes.
A pair of bumps (i.e., the tidal bulges)
are, then, formed at the disk inner edge,
as shown in panel (c) of Fig.~\ref{fig:snap_circ}.

\begin{figure*}
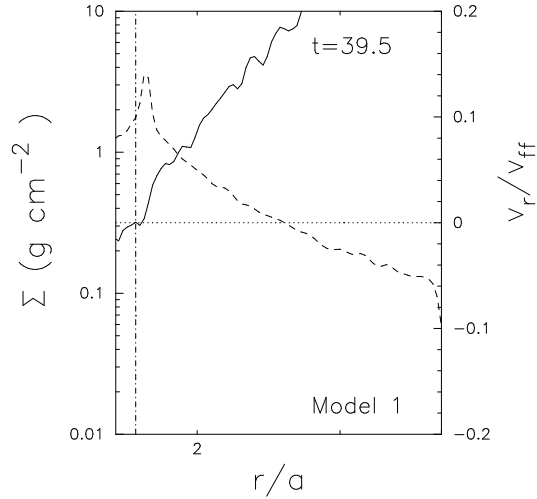

  \begin{center}
    \FigureFile(70mm,50mm){kh8.eps}
  \end{center}
 \caption{
          Radial distributions of the surface density (dashed line) and
          radial velocity (solid line) normalized by the free-fall velocity
          at $t=39.5$ in model~1.
          The vertical dash-dotted line indicates the position of
          the tidal truncation radius $r_{\rm{trunc}}=1.68 a$.
          A positive (or negative) radial velocity means
          the outward (inward) flow.
         }
 \label{fig:sdvr1}
\end{figure*}

Next, let us examine the radial structure of the circumbinary disk.
The left panel of Fig.~\ref{fig:sdvr1}
shows the radial distributions of the surface density and
radial velocity at $t=39.5$ in model~1.
The solid line, the dashed line and the dash-dotted line
show the surface density (in units of g~cm$^2$),
the radial velocity normalized by the free-fall velocity, and
the tidal truncation radius, $r_{\rm{trunc}}=1.68 a$, respectively.
A positive (or negative) radial velocity indicates an outward
(inward) flow.  Clearly,
the radial motion of the gas in the circumbinary disk
is mostly outward, whereas it is inward inside the
tidal truncation radius.
Note that the upwardly convex shape of the surface density distribution
may be due to an artifact of our way of mass injection
(recall that we inject mass to the radius of $r_{\rm{inj}}=1.73 a$).
However, it is unlikely that our treatment
will seriously influence the mass inflow rate
at the disk inner edge, since the circumbinary disk evolves
on much longer timescale than the orbital period.


\subsection{Mass Supply and Mass Capture}
\label{subsec:supply-capture}

\begin{figure*}
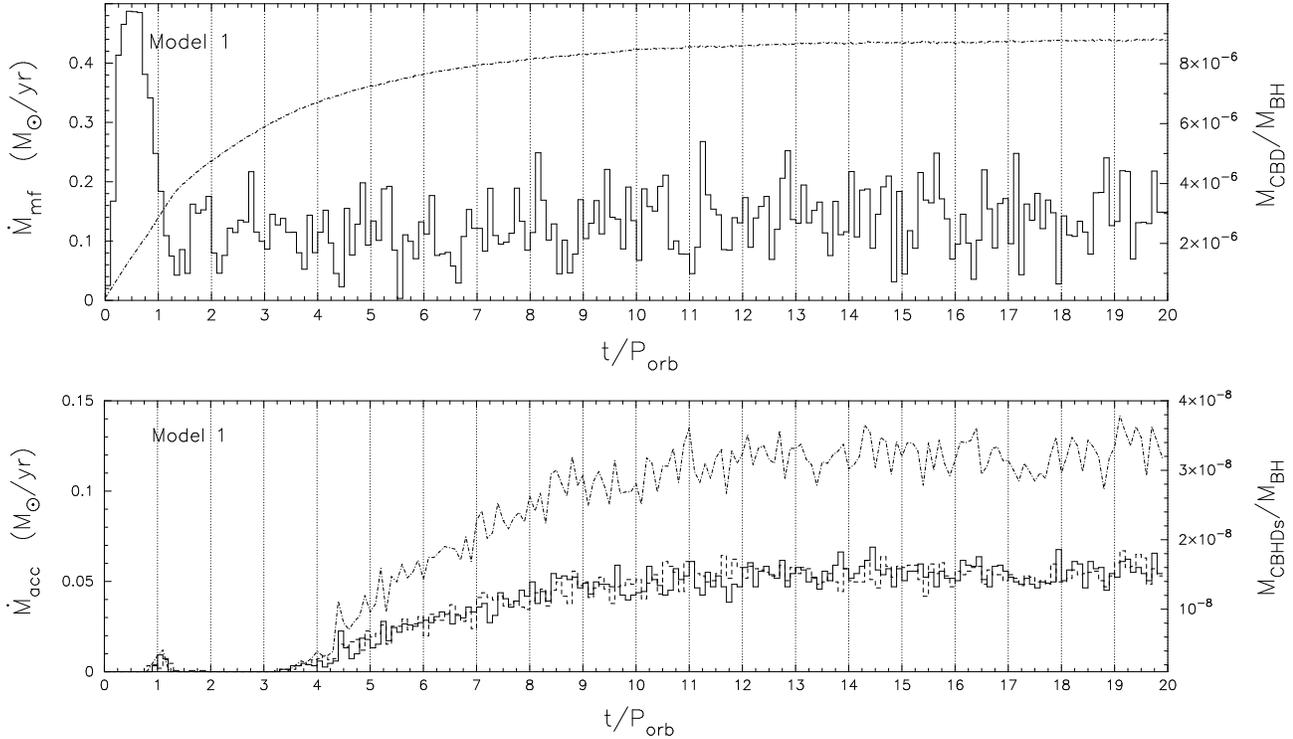

  \begin{center}
    \FigureFile(170mm,120mm){kh9.eps}
    \FigureFile(170mm,120mm){kh10.eps}
  \end{center}
 \caption{
Evolution of
the azimuthally averaged mass flux $\dot{M}_{\rm{mf}}$ at the tidal
truncation radius $r_{\rm{trunc}}=1.68 a$
and the mass of circumbinary disk $M_{\rm{CBD}}$ (upper panel),
and of the mass-capture rate $\dot{M}_{\rm{acc}}$
and the total mass captured by black holes
$M_{\rm{CBHDs}}$ (lower panel) in model~1.
In the upper panel,
the solid line and the dash-dotted line denote
the azimuthally averaged mass flux and the mass of the 
circumbinary disk 
normalized by total black-hole mass $M_{\rm{BH}}$, respectively.
In the lower panel,
the solid line, the dashed line 
and the dash-dotted line show the mass capture rate
of the primary black hole,
that of the secondary black hole 
and the total mass captured by black holes 
in units of $M_{\rm{BH}}$
, respectively.
          }
 \label{fig:mdotevo}
\end{figure*}

Let us next see long-term evolution of the
mass supply from the circumbinary disk
and the mass capture by the black holes.
Fig.~\ref{fig:mdotevo} illustrates the initial evolution
of the mass-supply rate (upper)
and the capture rates with (lower),
together with that of the circumbinary disk mass (upper) 
and of total mass captured by black holes (lower),
during $0\le{t}\le20$ in model~1.
The mass-capture rate is defined by how much the material 
is captured by
each black hole at the accretion radius $r_{\rm acc} = 0.2a$
in units of M$_{\odot}$ yr$^{-1}$.
The solid line, the dashed line and the dash-dotted line show
the mass-capture rates by the primary black hole
and by the secondary black hole, and the disk mass, respectively.
We see in this figure that
the disk mass steadily increases until $t=15$ and
then stays nearly constant afterwards.
Both mass-capture rates by the black holes also increase
until $t=15$ and saturate afterwards 
but show substantial fluctuations.
We can thus safely conclude that mass supply and accretion flow
reach their quasi-steady state after $t\simeq{15}$.

To investigate how the results depend on an initial disk-width,
we performed another simulation with the same simulation
parameters as those of model~1,
but with an initially more extended circumbiary disk;
the disk-inner edge is at 1.68a and the disk-outer edge is at 2.0a.
As a result, no appreciable changes have been seen
in the variation of the accretion rates with respect to the orbital
phase. We note, therefore, that the initial disk-width affect
little the dynamics of the mass inflow.


\subsection{Circumblack-hole Disk Formation}
\label{subsec:diskform}

\citet{kimi1} discussed the possibility
of the accretion disk formation around the
neutron star in a Be/X-ray binary using the data of SPH particles
captured by the neutron star.
In this subsection, we discuss a possibility of
the circumblack-hole disk formation around each black hole in model~1,
adopting a similar approach as that in section~2.3 of \citep{kimi1}.

\begin{figure}
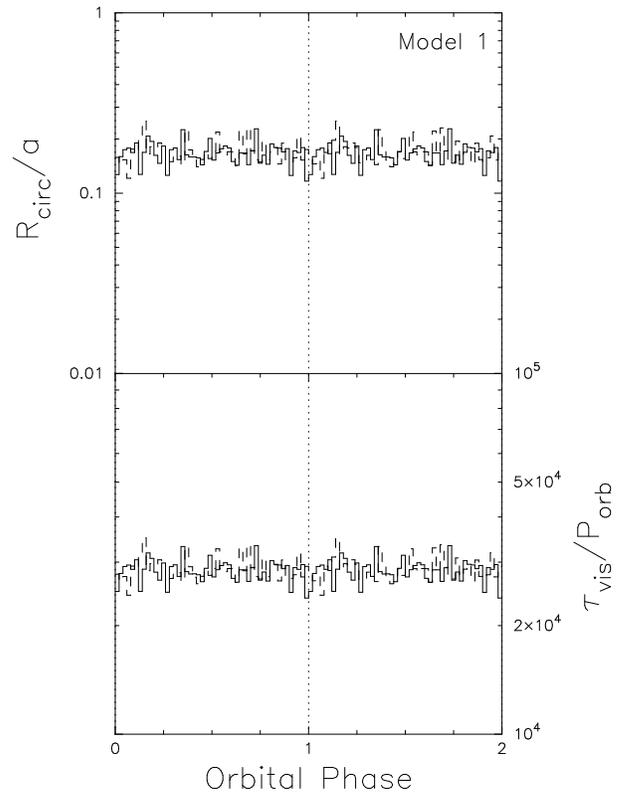

  \begin{center}
    \FigureFile(80mm,80mm){kh11.eps}
  \end{center}
\caption{
Orbital phase dependence of the circularization radius (upper panel)
and the ratio of the viscous-time scale to 
the orbital period (lower panel)
for the primary (solid lines) 
and the secondary (dotted lines) black hole,
respectively in model~1.}
 \label{fig:rc1}
\end{figure}

The material captured at $r_{\rm{acc}}$ has
a specific angular momentum $J$, by which we can infer
the circularization radius of the gas particles,
$R_{\rm{circ}}=J_{i}^{2}/GM_{i}$, where
suffix $i=$p refers to 
the primary black hole and $i=$s to the secondary one.
The upper panel of Fig~\ref{fig:rc1}
 shows the orbital dependence of the
circularization radius,
 where the solid line and the dashed line denote
the circularization radius 
of the primary and the secondary, respectively.
This figure clearly shows that
the circularization radii 
largely exceed the Schwarzschild radii
of the black holes.
Thus, the formation of a disk 
around each black hole is very likely.

The lower panel of Fig~\ref{fig:rc1} denotes
the viscous timescale of each circumblack-hole disk evaluated
at $R_{\rm{circ}}$.
For simplicity, we assume the circumblack-hole disk
to be geometrically thin and isothermal with the Shakura-Sunyaev
viscosity parameter $\alpha_{\rm{SS}}$.
The ratios of $\tau_{\rm{vis}}/P_{\rm{orb}}$ for the primary
and the secondary black holes are given, respectively, by
\begin{eqnarray}
\frac{\tau_{\rm{vis}}}{P_{\rm{orb}}}{\biggr|_{\rm{p}}}
&=&
\frac{1}{2\pi\alpha_{\rm{SS}}c_{\rm{s}}^{2}}
  \left(\frac{R_{\rm{circ}}}{a}\right)^{1/2}
\frac{GM_{\rm{p}}}{a}\left(1+q\right)^{1/2}, \\
\frac{\tau_{\rm{vis}}}{P_{\rm{orb}}}{\biggr|_{\rm{s}}}
&=&
\frac{1}{2\pi\alpha_{\rm{SS}}c_{\rm{s}}^{2}}
  \left(\frac{R_{\rm{circ}}}{a}\right)^{1/2}
\frac{GM_{\rm{s}}}{a}\left(1+\frac{1}{q}\right)^{1/2}.
\label{eq:ts}
\end{eqnarray}
The orbital phase dependence of $\tau_{\rm{vis}}/P_{\rm{orb}}$ is
shown in the lower panel of Fig~\ref{fig:rc1}.
It is immediately seen that the viscous timescales
in each of circumblack-holes disks
are much longer than the orbital period.


\section{Effects of Eccentricity and Unequal Masses}
\label{sec:uneq_ecc}

In this section, 
we first discuss the effects of an orbital eccentricity,
which gives rise to interesting orbital-phase modulations.
We then touch on the cases with unequal mass black holes.

\subsection{Eccentric Binary with Equal-Mass Black Holes }
\label{subsec:overview2}

The evolution of supermassive BBHs with
an orbital eccentricity
has been discussed (e.g. \cite{roos,PR,RT,QH,armi2}).
The eccentricity could grow secularly due to
the interaction between the black hole 
and its ambient stellar medium,
although this feature has 
not yet obtained general consensus.
In this section,
we describe the accretion flow from the circumbinary disk
onto the central binary with an eccentricity $e=0.5$.

\begin{figure*}
  \begin{center}
    \FigureFile(80mm,50mm){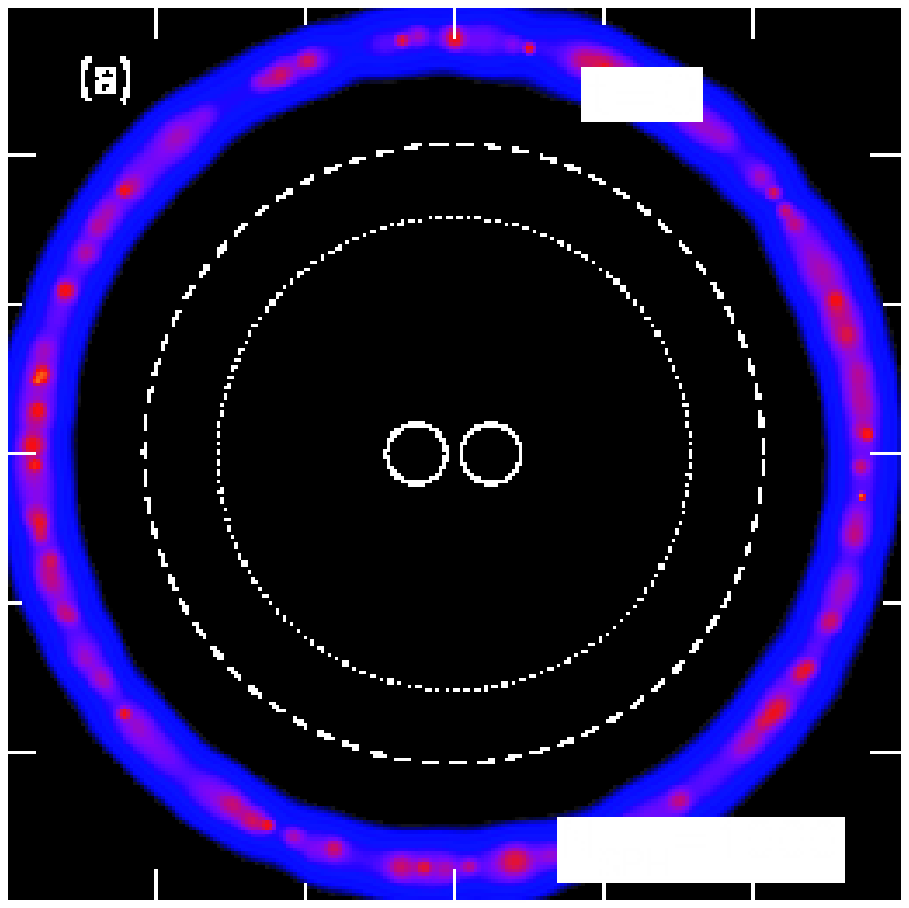}
    \FigureFile(80mm,50mm){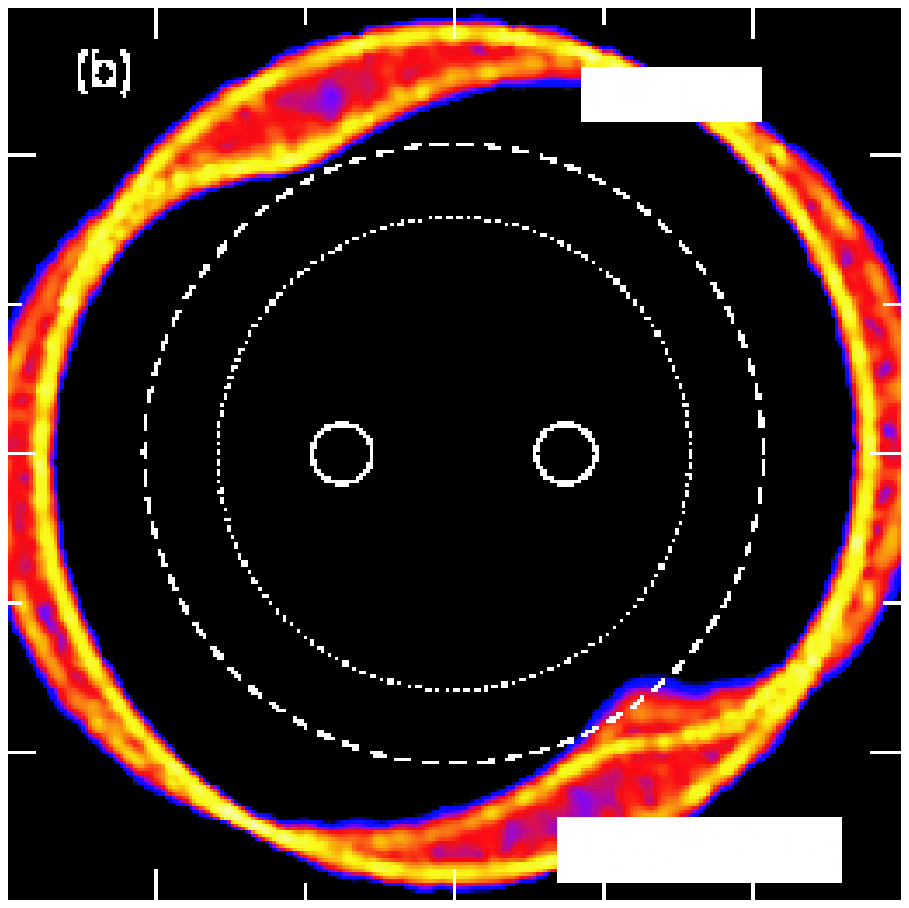}
  \end{center}
  \begin{center}
    \FigureFile(80mm,50mm){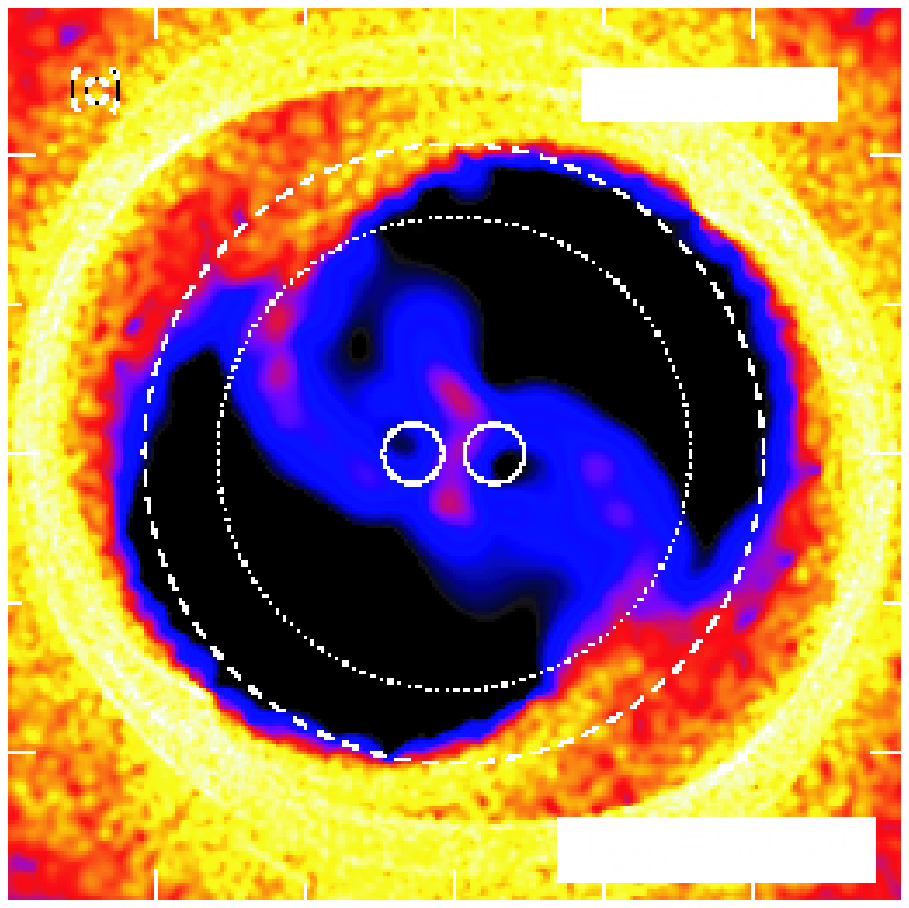}
    \FigureFile(80mm,50mm){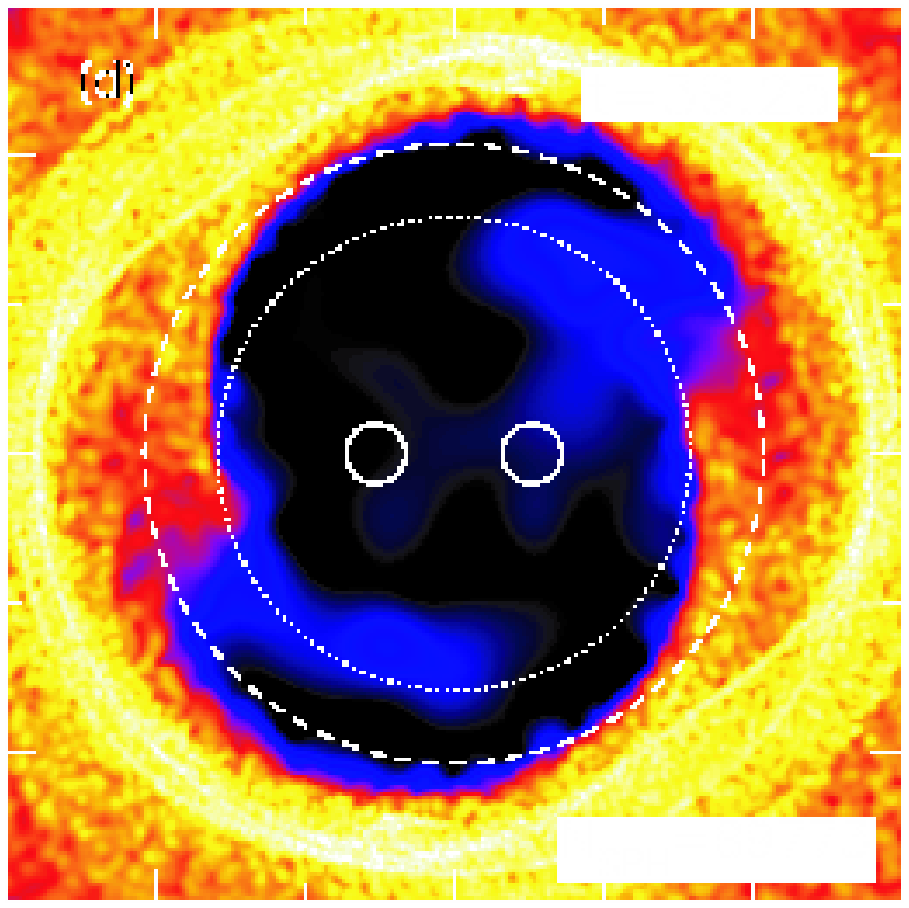}
  \end{center}
  \begin{center}
    \FigureFile(80mm,50mm){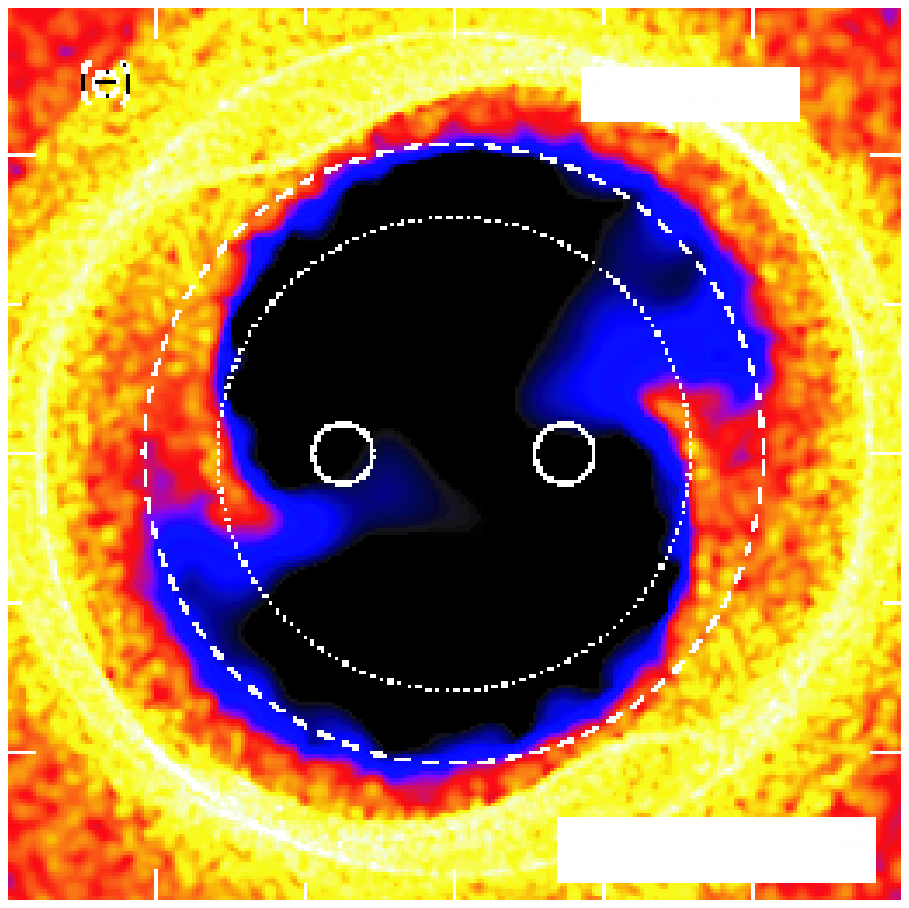}
    \FigureFile(80mm,50mm){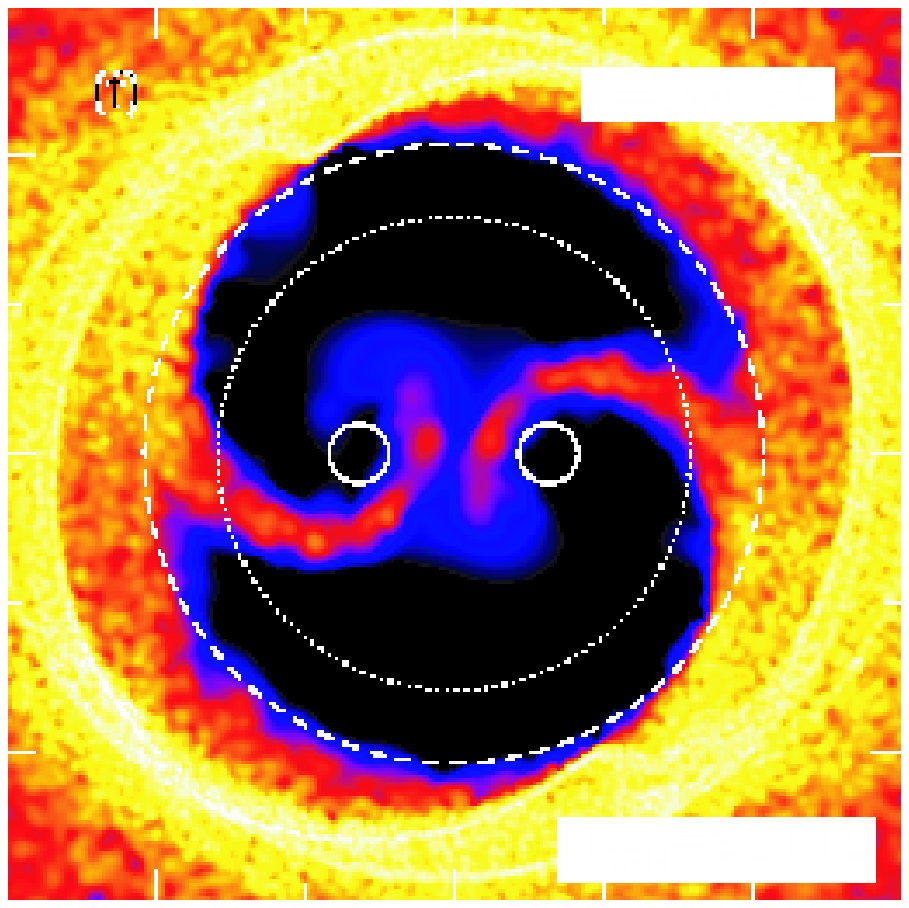}
  \end{center}
 \caption{
Same formats as model~1
(Fig.~\ref{fig:snap_circ}), but for model~2.
The dashed circle and the dotted circle denote
the (2,1) outer Lindblad resonance
radius and the (2,1) corotation radius, respectively.
         }
 \label{fig:snap_ecc}
\end{figure*}

Fig.\ref{fig:snap_ecc} shows
snapshots of accretion flow around
supermassive BBHs with eccentricity $e=0.5$ and equal
masses $q=1.0$.
Here, the dotted circle and the dashed circle correspond to
the radius of
the $(2,1)$ corotation resonance
and that of
the $(2,1)$ outer Lindblad resonance, respectively,
where the $(m,l)$ corotation resonance radius and
the $(m,l)$ outer Lindblad resonance radius 
are given, respectively, by
$(m/l)^{2/3}a$ and $((m+1)/l)^{2/3}a$ 
for a circumbinary disk.
Here $m$ and $l$ are the azimuthal 
and time-harmonic numbers, respectively,
in a double Fourier decomposition of the binary potential,
$\Phi(r,\theta,t)=
\sum\phi_{m,l}(r)\exp[i(m\theta-l\Omega_{\rm{B}}t)]$, where
$\Omega_{\rm{B}}$ is 
the orbital frequency of the binary \citep{arty3}.

After the disk is set up (see panel (a) of Fig.~\ref{fig:snap_ecc}),
the tidal bulge is formed and the gas reaches
the radius of
the (2,1) outer Lindblad resonance at $t=1.5$,
as is seen in panel (b) of Fig~\ref{fig:snap_ecc}.
The semi-major axis of the elongated disk
is not aligned with that of the binary,
as well as in a circular binary (see panel (b) of Fig.~\ref{fig:snap_circ}).
This phase shift of the tidal bulge occurs by the resonance friction.
The unique feature seen in the case of eccentric binaries is that
since the binary separation is periodically
changing with time,
the phase shift should also be changing with time.
In addition, the gravitational attraction force 
to the circumbinary disk
also changes periodically
and is maximum for the closest parts of the disk to black holes
at the phase of the maximum binary separation
(i.e., at the apastron).

The material around the binary inside $r_{\rm{trun}}$ 
will be swept away,
since it acquires angular momentum transferred from the BBHs.
This transfer is induced by the
resonance interaction between the binary and the circumbinary disk
(see \authorcite{arty3} \yearcite{arty3,arty4,arty5};
\authorcite{lubow1} \yearcite{lubow1,lubow2}).
As a result,
the inner edge of the circumbinary disk is truncated
roughly at the (2,1) outer Lindblad resonance
as seen in panels (c)-(f) of Fig~\ref{fig:snap_ecc}.
This supports the results of \citet{arty4}.


\subsection{Gap Opening and Closing}

Panels (c)-(f) of Fig~\ref{fig:snap_ecc} represent
sequential snapshots of accretion flows in this system
for $39<{t}<40$.
While the material inside the (2,1) corotation radius
is captured by the black holes,
the mass outside the (2,1) corotation 
radius outwardly flows at $t=39.02$.
Subsequently, most of the material is swallowed by black holes
at $t=39.25$, whereas the gas keeps the outward flow outside
the (2,1) corotation radius.
The mass overflow is, then, initiated at $t=39.5$,
and the gas falls and reaches the BBHs at $t=39.79$.
The most conspicuous feature in the evolution of model~2
resides in this periodic on/off transitions of mass supply.
In the absence of the mass inflow from the circumbinary disk
a big gap appears between the disk and the black holes.
We call this the gap-opening phase.
When the mass inflow occurs from the circumbinary disk
and then make bridges from the disk to the BBHs,
the gap is closed.  This is the gap-closing phase.
The typical four stages in one gap-closing and opening cycle is repeated
in sequential order.
In model~1, in contrast,
there is a continuous mass supply to the black holes
(i.e., the gap is always closed).

\begin{figure*}
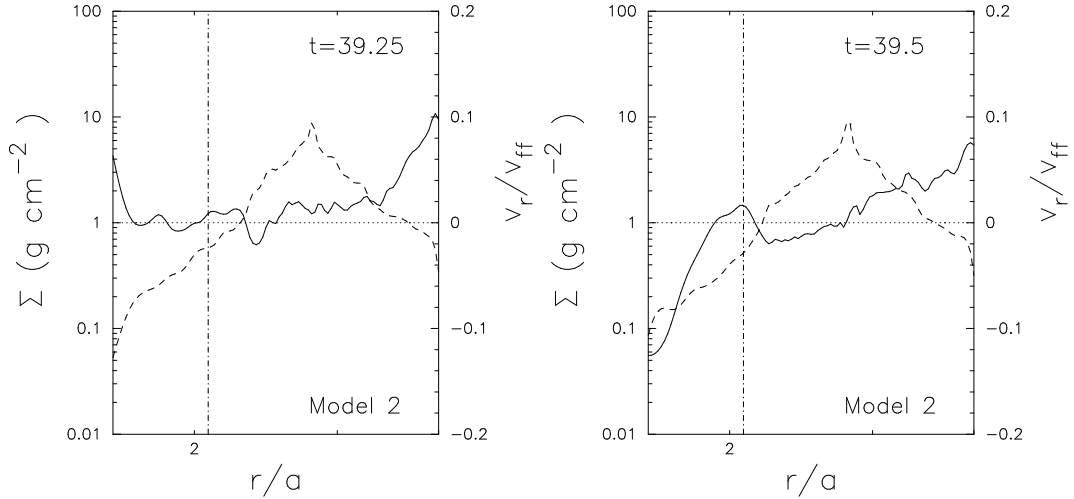

  \begin{center}
    \FigureFile(70mm,50mm){kh18.eps}
    \FigureFile(70mm,50mm){kh19.eps}
  \end{center}
 \caption{
Same format as model~1 
(the left panel of Fig.~\ref{fig:sdvr1}), but for model~2.
The results are shown at two different phases:
at phase $0.25$ and $0.5$ in the left and the right panels,
respectively.
The vertical dash-dotted line shows
the $(2,1)$ outer Lindblad resonance radius.
         }
 \label{fig:sdvr2}
\end{figure*}

As seen in panel~(c) of Fig.~\ref{fig:snap_ecc},
the circumbinary disk stops mass inflow towards its inner edge
when the binary is at the periastron (phase 0),
whereas the mass which already left the (2,1) corotation radius
continues to fall on to each black hole 
by the gravitational attraction.
Here, Fig.~\ref{fig:sdvr2} displays the radial distributions
of the surface density and the radial velocity in the circumbinary disks
in the two different phases.
The left panel shows that the radial velocity is almost everywhere
outward at phase 0.25, meaning no mass inflow,
whereas there is mass inflow at the apastron (phase 0.5).

We need to distinguish the following two steps to understand the
orbital modulation of the mass flow stream:
(1) When the binary is at the periastron,
the angular momentum is much more transferred from the binary
to the circumbinary disk than otherwise.
Accordingly, the gas inflow is terminated after the periastron
[see panel (c)].
Conversely,
the mass inflow is at maximum around the apastron.
(2)
the material which was launched from the circumbinary disk
at the apastron
will take some time to reach the BBHs.
Hence, there arises a phase-lag between the mass-supply maximum
and the mass capture maximum,
which roughly corresponds to the free-fall
time from the circumbinary disk to the black holes.
As a result, the majority of infalling gas is captured by the black holes
just before the periastron [see panel (f)].
Hence, the gap starts to be closed after the apastron and
is completely closed before the periastron (at phase $\sim$ 0.75),
and then is opened again after the periastron (at phase $\sim$ 0.25).


\subsection{Mass Supply and Mass Capture}


\begin{figure*}
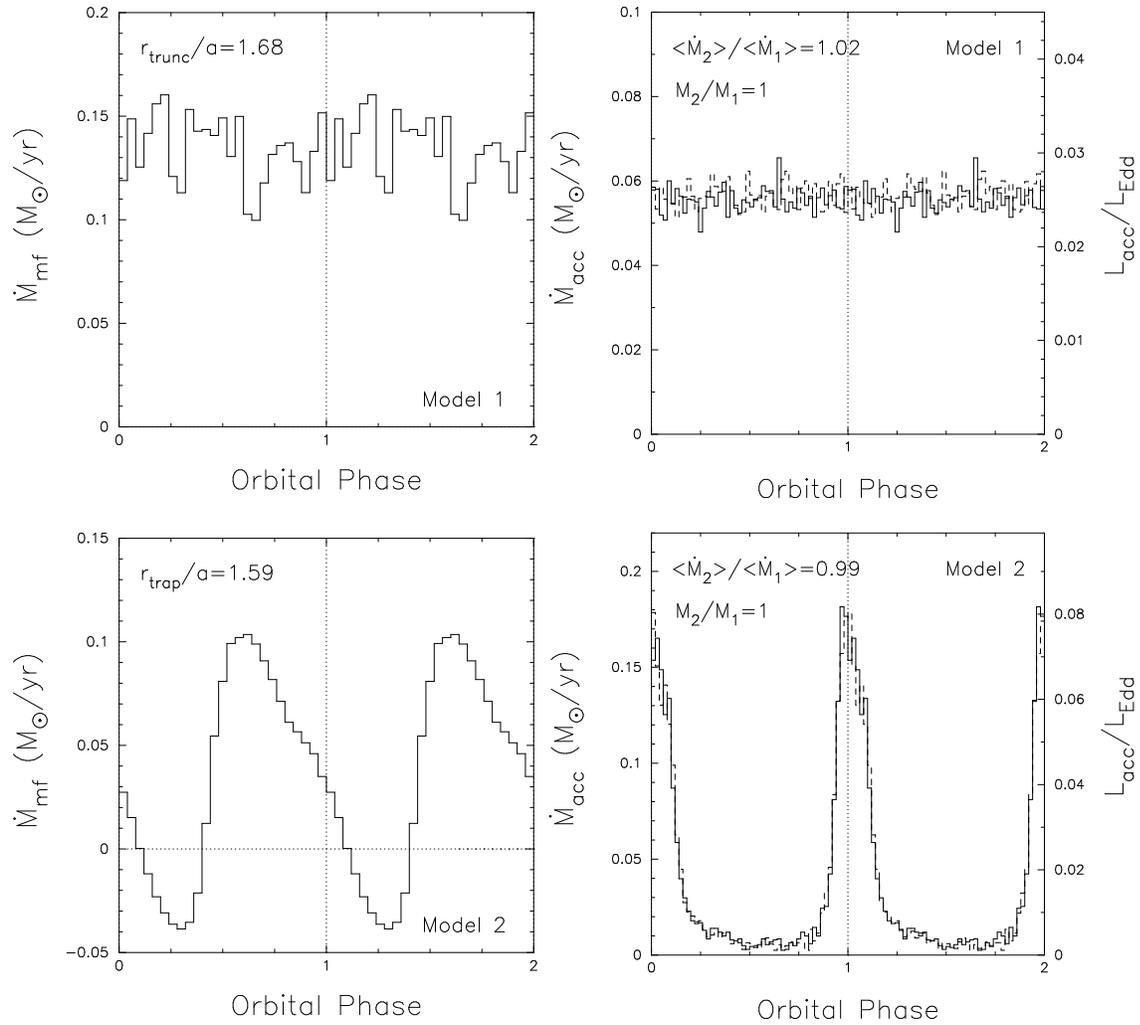

  \begin{center}
    \FigureFile(70mm,50mm){kh20.eps}
    \FigureFile(78mm,50mm){kh21.eps}
  \end{center}
  \begin{center}
    \FigureFile(70mm,50mm){kh22.eps}
    \FigureFile(78mm,50mm){kh23.eps}
  \end{center}
 \caption{
Orbital-phase dependence of azimuthally averaged mass flux
and mass-capture rate in models~1 (upper panels) and 2
(lower panels), respectively.
The binary is at the periastron (apastron) at phase 0.0 (0.5).
The data are folded on the orbital period
over $20\le{t}\le40$ in model~1 and over $40\le{t}\le60$ in model~2.
The mass flux is measured at
the tidal truncation radius $r_{\rm{trunc}}=1.68 a$
in model~1 and at the (2,1) corotation
resonance radius $r_{\rm{trap}}\simeq1.59 a$ in model~2, respectively.
The right axis shows the bolometric luminosity corresponding to the
mass-capture rate with the energy conversion efficiency, $\eta=0.1$,
normalized by the Eddington luminosity with total black hole mass
$M_{\rm{bh}}=1.0\times10^{8}M_{\odot}$,
where $\eta$ is defined by 
$L_{\rm{acc}}=\eta\dot{M}_{\rm{cap}}c^{2}$.
         }
 \label{fig:opd_m1m2}
\end{figure*}

Owing to the reason mentioned above,
there arises a phase delay between
the moment of the maximum azimuthally-averaged mass flux and
that of the maximum mass-capture rate.
We next show how the mass-capture rate and
the averaged mass flux vary with binary orbital motion.
To reduce the fluctuation noise, 
these data are folded on the orbital period over $40\le{t}\le60$
after the system is the quasi-steady state ($t\ge38$).

The lower-left panel of Fig.~\ref{fig:opd_m1m2} shows
the azimuthally averaged mass flux at the (2,1) corotation radius
in model~2.
While the circumbinary disk has an outward flow from 
the periastron passage to the phase somehow before apastron,
the mass is inwardly launched from the disk-inner edge 
from the phase somehow before apastron to the next periastron.
This also supports that the gap is opening after the periastron
and the gap is closing after the apastron in the eccentric binary.

The lower-right panel of Fig~\ref{fig:opd_m1m2}
represents the orbital-phase dependence
of the mass-capture rate
and the corresponding luminosity normalized Eddington luminosity
with total black hole mass $10^{8}\MO$;
i.e., $L_{\rm E}\sim 10^{46}$ erg s$^{-1}$, in model~2.
Despite the fact that we continuously inject mass
to the circumbinary disk at a rate of
$\dot{M}_{\rm{inj}}=1.0\,M_{\odot}\rm{yr}^{-1}$,
which would produce about the Eddington luminosity for
$\eta\sim 0.1$,
the calculated luminosity is substantially sub-Eddington.
This is because the majority of the injected mass is lost in this system.
The solid line and dashed line
denote the mass-capture rate by the primary black hole
and by the secondary one, respectively.
This figure clearly shows that
the mass-capture rate
significantly modulates with the orbital motion.
The peak of mass-capture rate is located in 
before the periastron passage.
Furthermore, 
the ratio of the lowest mass-capture rate to the highest one
during one orbital period is $\sim1/10$.

In the context of the young binary star formation,
a couple of simulations with the similar mass ratio and orbital
eccentricity as those of model~2
were performed by
\cite{arty4} and \cite{GK1}.
They are
the non-dimensional, high-eccentricity system
with (q,e)=(1,27,0.5) by \cite{arty4} and
AK Sco system with (q,e)=(0.987$\pm$0.007,0.469$\pm$0.004)
by \cite{GK1}, respectively.
The orbital modulation of mass accretion rates in both cases
are well corresponding to that of model~2.
The ratios of the lowest mass accretion rate to the
highest one are the factor 7-8, which is lower
than that of model~2 because of their smaller accretion radii.

For comparison purpose,
we give the same but for model 1
in the upper panels of Fig.~\ref{fig:opd_m1m2}.
This figure clearly represents that
the circumbinary disk continues to supply
the gas to its inner regions
over the whole orbital period.
Then, the gas falls onto
each black hole with fluctuations.
This behavior is mainly caused by the overflow
as seen in the last two panels of Fig~\ref{fig:snap_circ}.

\subsection{Unequal Masses}
\label{subsec:uneq}

\begin{figure*}
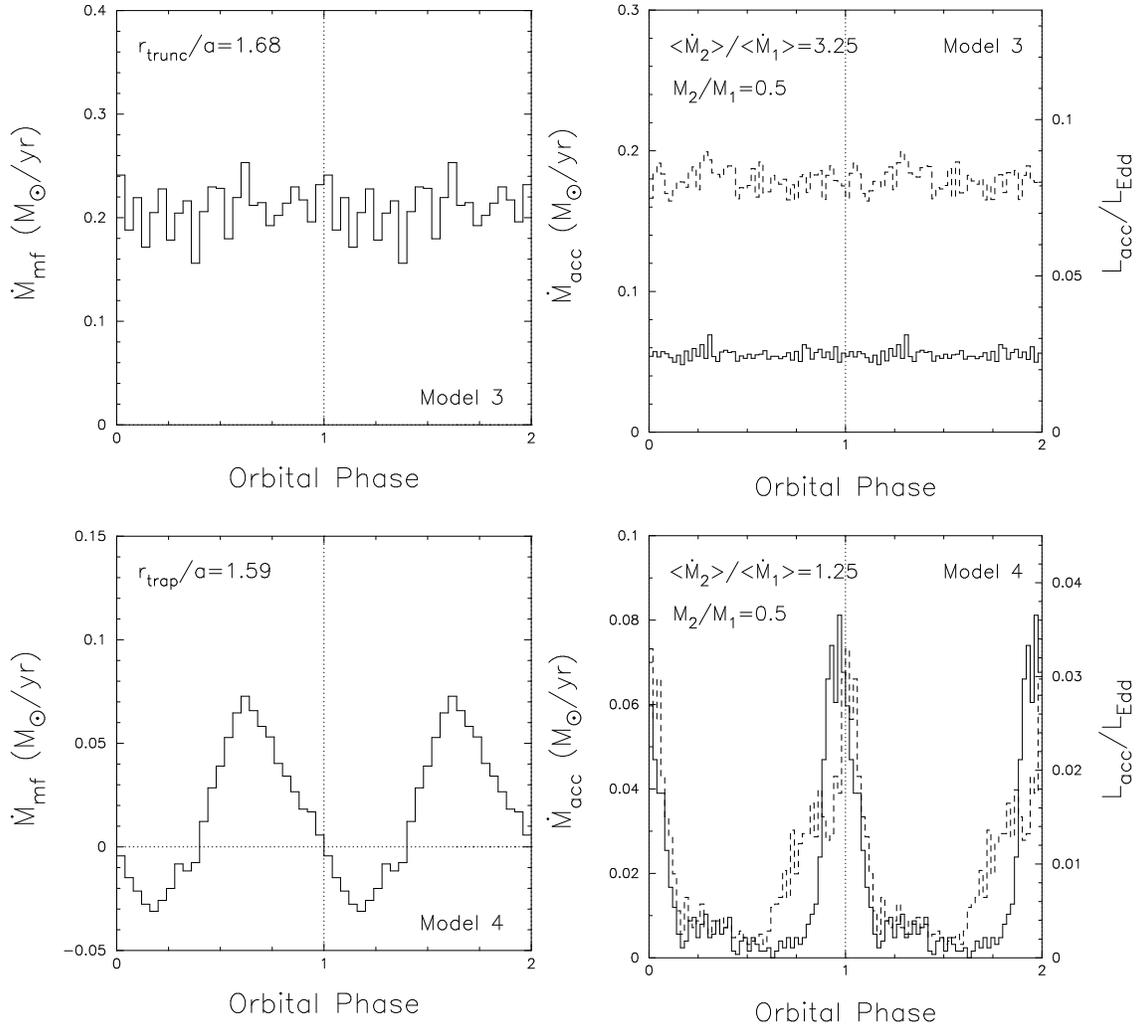

  \begin{center}
    \FigureFile(70mm,50mm){kh24.eps}
    \FigureFile(78mm,50mm){kh25.eps}
  \end{center}
  \begin{center}
    \FigureFile(70mm,50mm){kh26.eps}
    \FigureFile(78mm,50mm){kh27.eps}
  \end{center}
 \caption{
Same formats as model~1, but for 
models~3 (upper panels) and 4 (lower panels), respectively.   
         }
 \label{fig:uneq}
\end{figure*}

Finally, we discuss the cases of unequal mass black holes (models~3 and 4).
The most common binaries are likely to possess
black holes with extreme mass ratios
in the case of the minor galactic mergers
(\cite{armi1,armi2}).
If the binaries are formed as a result of a major merger,
on the contrary, its mass ratio
can be non-extreme and slightly deviate from unity. 
The effects of non-extreme but
unequal masses on the accretion flow onto the supermassive BBHs
are investigated by models~3 and 4 and
are remarkably represented by
the azimuthally averaged mass flux and
the mass-capture rate by each black hole
in Fig.\ref{fig:uneq}.

Model~3 is the run with $(q,e)=(0.5,0.0)$.
It is seen from the upper panels of Fig.\ref{fig:uneq} that
the mass-capture rate of the secondary black hole
is much higher than that of the primary one.
This is because the distance between the 
inner edge of the circumbinary disk and the $L_{2}$ point
is shorter than that between the disk inner edge and
$L_{3}$ point. Thus, the gas prefers to be attracted
from the secondary black hole 
closer to the $L_{2}$ point than the primary one.
Here, note that the total 
mass-capture rate by both of black holes roughly 
equals to the azimuthally averaged mass flux.

In model~4, which is run with $(q,e)=(0.5,0.5)$,
the lower panels 
of Fig.\ref{fig:uneq} indicates that 
the rising time of the secondary burst is earlier 
than that of the primary burst 
( burst in gas accretion onto the primary black hole). 
Thus, the amount of gas captured by the secondary black hole
during one orbital period is more than that by the primary
black hole.
This is consistent with the results of model~3, i.e.,
the gas is easier to be captured by the secondary black hole.


\section{Discussion}
\label{sec:discussion}

We have performed the SPH simulations of
accretion flow around the supermassive BBHs.
We find that the material overflows from the circumbinary disk
via two points, freely infalls towards either of the BBHs,
and is eventually captured by it.
While the mass capture-rate
has little orbital-phase dependence in the case of a circular binary,
it exhibits significant orbital modulations in the case of an eccentric binary.
In this section, we discuss
the formation and evolution processes
of the circumblack-hole disks, circumbinary disk evolution,
shock formation near the circumblack-hole disks, and
possible strategy to detect such BBH candidates
exhibiting periodic light variations.

\begin{table*}
\begin{center}
\caption{
Summary of simulation results
The first column represents model numbers.
The second column and the third column 
are the mean accretion rates of the primary black hole and
those of the secondary black hole,respectively.
The fourth column and the fifth column are 
the mean circularization radii of the primary and
those of the secondary,respectively.
The last two columns are 
the mean viscous time-scales of the primary black hole and of the
secondary black hole,respectively.
All quantities shown in the 2-7th columns 
are folded on the orbital period over
$20\le{t}\le40$ in modes~1 and 3, 
and $40\le{t}\le60$ in models~2 and 4.
}
\label{tbl:results}
\begin{tabular}{@{}lccccccc}
\hline
Model       & $<\dot{M}_{\rm{p}}>$
            & $<\dot{M}_{\rm{s}}>$
            & $<R_{\rm{circ},\rm{p}}>$
            & $<R_{\rm{circ},\rm{s}}>$ 
            & $<t_{\rm{vis},\rm{p}}>$
            & $<t_{\rm{vis},\rm{s}}>$ \\
            & $(M_{\odot}\rm{yr}^{-1})$
            & $(M_{\odot}\rm{yr}^{-1})$
            & $a$
            & $a$ 
            & $P_{\rm{orb}}$
            & $P_{\rm{orb}}$ \\
\hline
1           & $5.6\times10^{-2}$
            & $5.7\times10^{-2}$
            & $1.5\times10^{-1}$
            & $1.5\times10^{-1}$ 
            & $2.7\times10^{4}$ 
            & $2.7\times10^{4}$ \\
2           & $3.8\times10^{-2}$
            & $3.8\times10^{-2}$
            & $8.1\times10^{-2}$
            & $8.5\times10^{-2}$ 
            & $2.0\times10^{4}$ 
            & $2.0\times10^{4}$ \\
3           & $5.5\times10^{-2}$
            & $1.8\times10^{-1}$
            & $1.0\times10^{-1}$
            & $4.2\times10^{-1}$ 
            & $2.6\times10^{4}$ 
            & $3.7\times10^{4}$ \\
4           & $1.7\times10^{-2}$
            & $2.1\times10^{-2}$
            & $7.7\times10^{-2}$
            & $8.4\times10^{-2}$ 
            & $2.1\times10^{4}$
            & $1.6\times10^{4}$ \\
\hline
\end{tabular}
\end{center}
\end{table*}

\subsection{Circumblack-hole Disks}
\label{subsec:cbhdisks}

We have already discussed
in section~\ref{subsec:diskform}
a possibility of the formation of
the circumblack-hole disks and estimated
their viscous timescales at the circularization radii.
In this section,
we discuss whether the circumblack-hole disks
are persistent or not during one orbital period
and how viscous accretion processes affect
the evolution and structure of the circumblack-hole disk.

Table~\ref{tbl:results}
summarizes the mean mass-capture rates,
the mean circularization radii, and
the mean viscous timescales of all the calculated models.
To reduce the fluctuation noise,
the simulation data are folded on the orbital period
over $20\le{t}\le40$ in models~1 and 3,
and over $40\le{t}\le60$ in models~2 and 4.
The second and the third columns
denote the mean mass-capture rates of the primary
and the secondary black holes, respectively.
These columns clearly show that
the sum of the mean mass-capture rates by the BBHs
in the circular binaries
is higher than that in the eccentric binaries.
This implies that
the circumblack-hole disks in the circular binaries
are denser than those in the eccentric binaries.

We see from
the fourth column and the fifth column of Table~\ref{tbl:results}, which
denote the mean circularization radii
around each black hole, that
the size of the circumblack-hole disks in the circular binaries is
larger than that in the eccentric binaries.
In model~3,
the disk size of the secondary black hole, $R_{\rm{circ,s}}=0.42a$,
is larger than the accretion radius, $r_{\rm{acc}}=0.2a$.
This means that some of the mass
which enters the sphere of $r_{\rm{acc}}=0.2a$ may finally go out,
without being captured by the black hole.

For comparison with model~1, we've performed the other simulation model
with the initially more extended circumbinary disk;
the disk-inner edge 1.68a and the disk-outer edge
2.0a where the mass is injected (see Section~\ref{subsec:supply-capture}).
Although the circularization radii are less than half those of model~1,
our conclusion regarding the formation of circumblack-hole disks is 
unchanged.

The last two columns of Tabel~\ref{tbl:results}
represent the mean viscous time-scale of
the primary black hole and of the secondary one, respectively.
These are estimated by using equation~\ref{eq:ts} and \ref{eq:ts}.
It is clear from these columns that the
viscous time-scale is much longer than the orbital
period in all the models.
Thus, the circumblack-hole disks
once formed, 
they will survive over the whole orbital phase. 
The viscosity in the circumblack-hole disks
gives little influence on their structure and 
short-term evolution 
on the timescale of the order of the orbital period.

The phase-dependent accretion flows are likely to give a good impact
on the outer edge of the circumblack-hole disks.
This will be able to excite
one-armed oscillations on the disk (e.g., \cite{kimi2}).
As the one-armed waves propagates, the material
on the outer region of the disk is inwardly pushed towards
the central black hole.
Since the propagation time-scale 
is roughly estimated 
by using $(\alpha_{\rm{SS}}/2\pi)\tau_{\rm{visc}}$,
where $\tau_{\rm{vis}}$ is 
the viscous time-scale of circumblack-hole disks (see \cite{kato}),
as $\sim10^{2-3}P_{\rm{orb}}$,
the outer region of the disk can significantly vary,
whereas the inner region of the disk may remain to be unchanged.
We thus expect that optical/IR radiation emitted mainly from the outer portions
will exhibit significant periodic variations, whereas
radio and X-ray emissions coming from the innermost region
may not show such periodic and coherent variations
(however, see the later Section~\ref{subsec:shock}).
If this is the case,
the spectral energy distributions (SEDs) will be highly time-variable
on the orbital timescale.

It is likely that the mass
exchanges occur via an effective L1 point
between the disks around the primary 
and secondary black holes ( e.g. \cite{GK2}).
In addition to the phase-dependent mass inflow
from the circumbinary disk,
this mass exchange 
could also affect the evolution 
and structure of circumblack-hole disks.
This effect cannot be treated in our simulations, however,
because of the lack of sufficient resolution
in the narrow region around the effective L1 point.

The right-upper (lower) panels of 
Fig.~\ref{fig:opd_m1m2} and Fig.~\ref{fig:uneq},
also demonstrate that
the mass-capture rate is about one-order of magnitude
as low as the mass-input rate in all the models.
In fact, the total mass of circumblack-hole disks,
which approximately equals to the total mass captured by
black holes $M_{\rm{CBHDs}}$, is
much lower than the mass of circumbinary disk, as shown in 
Fig.~\ref{fig:mdotevo}.
This strongly suggests that the circumblack-hole disk has a significantly
low density. 
Thus, we expect that the accretion flow 
should become radiatively inefficient 
in the vicinity
of black holes \citep{kato}.

\subsection{Circumbinary Disks}
\label{subsec:cbdisks}

We address a question; what happens if the circumbinary disk
is inclined from the orbital plane of the supermassive BBHs?
The mass-capture rate profile may show
a two-peaked feature, because there are a couple of points
in the disk-inner edge to which the BBHs approach
during one orbital period.
In fact, the double-peak structure is observed in the optical
outbursts of OJ287 \citep{SSI}.
The effect of the inclination angle on the mass-capture rate profile
will be also examined in
a subsequent paper.

\citet{mac} asserted that the mass supply rate from the circumbinary disk 
to the BBHs exhibits a quasi-periodic modulation
due to the eccentricity of the circumbinary disk,
even if the binary black hole has circular orbit.
The disk eccentricity is excited 
due to the resonance interaction between the circumbinary disk 
and the central binary after
a few viscous time-scale of the circumbinary disk.
Such a long-term evolution of the circumbinary disk 
could also give
the time variations of the light curve 
of circumbinary disk itself, such as 
a superhump phenomenon in Dwarf Novae systems 
(e.g., \cite{Murray}).

The circumbinary disk evolution could play
a key role to resolve the loss-cone problem (e.g., \cite{arty1,mac}).
The disk-binary interaction gives an influence on
the global evolution of the binary orbital elements, e.g., the eccentricity. 
the semi-major axis, the mass-ratio, the inclination angle 
and so on \citep{arty2,bate1,lubow2}.
In the framework of the disk-binary interaction, therefore,
the effects of all the orbital elements
should ideally be taken account of. 

In addition,
it is an open problem how the gas is supplied
from an outer region over several parsec
to an inner, subparsec region,
in the context of the supermassive BBH systems 
in merged galactic nuclei.
The mass supply to the circumbinary disk
may prevent the accretion rate
from decreasing due to the viscous diffusion of the surface density.
Although this viscous diffusion could be neglected in such a short-term
evolution as our simulations,
it should be taken account of in the mass inflow processes
in the long-term evolution of the circumbinary disk 
over the viscous time-scale.
The long-term evolution of the circumbinary disk 
remains as a matter to be discussed further.

\subsection{Shock Formation}
\label{subsec:shock}

\begin{figure}
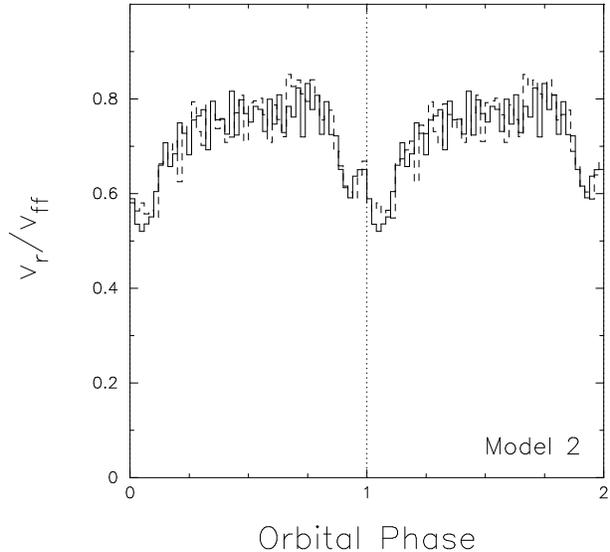

  \begin{center}
    \FigureFile(80mm,80mm){kh28.eps}
  \end{center}
 \caption{
Orbital dependence of the radial velocity 
normalized by the free-fall velocity 
at the inner boundary, $r_{\rm{acc}}=0.2a$.
         }
 \label{fig:vrvff}
\end{figure}

Another important issue to be considered is a possible formation of shock
structure.  We measured
the radial velocity at the inner boundary $r_{\rm{acc}}=0.2a$
in model~2
and plot its orbital-phase dependence in Fig.~\ref{fig:vrvff}.
Here,
the radial velocity is normalized by the free-fall velocity.
We understand that the radial
velocity is on the same order of magnitude
of the free-fall velocity and that it suffers the orbital modulation
with the peak being around the apastron (phase 0.5).
This trend contrasts with that of
the mass-capture rate which exhibits a peak around the periastron
(see Fig. \ref{fig:opd_m1m2}).
If the material infalls near circumblack-hole disk,
shock structures
could be formed near the outer edge of the circumblack-hole disks.
Substantial fraction of the kinetic energy of the material will be converted
to the thermal energy.
Consequently, detectable soft X-rays or UV could be periodically
emitted from the shock structures.
Such X-rays or UV will exhibit a broad peak around phase $\sim 0.5$, whereas
optical/IR radiation will show a sharp peak around phase $\sim$ 0.0.


\subsection{Observation Implications}
\label{subsec:observations}

\citet{sill} firstly proposed that 
the periodic behavior of the optical light-curve of OJ287 
may be induced by the orbital motion of the binary black holes,
that is,
the tidal perturbation of the secondary black hole
causes the accretion rate
of the primary black hole to be enhanced.
Although a number of models 
has been proposed during the last two decades since then
\citep{LV,katz,valtao},
no widely accepted mechanism has been yet proposed for
the behavior of periodic outbursts in OJ287.
Furthermore, 
several quasars are known often to show periodic
outbursts similar to those in OJ287. 
The existence of supermassive BBHs 
has been also suggested 
for these sources, 
for example, 3C75 \citep{YI}, 3C279 \citep{AC}, 
PKS 0420-014 \citep{britzen}
and 3C345 \citep{LR}.
Such periodic behaviors 
in the observed light curves of these sources
could be also explained by our scenario
i.e., the outbursts are driven 
by the orbital eccentricity of BBHs in three-disk systems, 
as shown in the
lower right panel of Fig.~\ref{fig:opd_m1m2} 
and Fig.~\ref{fig:uneq}.

Multiwavelength long-term monitoring observations 
should be a powerful tool to probe the existence
of the circumblack-hole disks, as well as that of supermassive BBHs. 
As described in Section~\ref{subsec:cbhdisks},
the photons emerging from the inner region of the disk
exhibits different SEDs and time variations from 
those emerging from the outer region.
In addition, it is expected that 
the shock-induced radiation should also show
periodic variations at different phases from those of SEDs, 
as seen in Fig.~\ref{fig:vrvff}.
If we detect periodic time variations of SEDs with
the different behavior at different wavelengths,
it provides the strong observational evidence for
the existence of supermassive BBHs in three-disk systems
on a parsec/subparsec scale
of the galactic center.

Finally,
we give a remark on the quasi-periodicities in the light curves.
These are often found in the blazar light curves
(see, e.g., \cite{kata}).
In this context, it is interesting to note that
\citet{nego}
have found quasi-periodic light variations
through the analysis of X-ray intensity variations from Cygnus X-1.
It is well known that its X-ray curves are composed of
numerous shots (or mini-flares) with a variety of flare amplitudes.
By picking up only large shots they found that the light variations
follow log-normal distribution.  That is,
if only large flaring events in the accretion flow produce
blob outflow in the form of a jet, jet light curves will be quasi-periodic.
This gives another possibility to produce quasi-periodicities
in blazar light curves. 
Much more work is desirable 
both theoretically and observationally
in this area.

\section{Conclusions}
\label{sec:conclusions}

For the purpose of
providing the observable diagnosis to probe
the existence of supermassive BBHs 
on a subparsec scale in merged galactic nuclei,
we have carried out the SPH simulations 
of accretion flows from circumbinary disks onto the supermassive BBHs. 
Our main conclusions are summarized as follows:

\begin{enumerate}
\renewcommand{\theenumi}{(\arabic{enumi})}
\item 
There are two-stage mechanisms to cause an accretion flow
from the circumbinary disk onto the supermassive BBHs:
First, 
the gas is guided to the two closest points on 
the circumbinary disk from the black holes
by the tidal deformation of the circumbinary disk.
Then, the gas is increasingly accumulated on these two points
by the gravitational attraction of the black holes.
Second, when the gas can pass
across the maximum loci of the binary potential,
the gas overflows via these two points
and inspirals onto the black holes 
with the nearly free-fall velocity
\item 
In circular binaries, the gas continues to be supplied
from the circumbinary disk onto the supermassive BBHs 
(i.e. the gap is always closed).
\item 
In eccentric binaries, 
the material supply undergoes the periodic on/off transitions
with an orbital period because of
the periodic variations of the binary potential.
\item 
While the mass-capture rates exhibit 
little orbital-phase dependence in circular binaries,
they significantly modulate with the orbital phase 
due to the gap opening/closing cycles,
in eccentric binaries.
This could provide the observable diagnosis 
for the presence of supermassive BBHs in three-disk systems
at the galactic center.
\item 
The circumblack-hole disks are formed around each black hole
regardless of the orbital eccentricity and the mass ratio.
\end{enumerate}

\bigskip

Authors are grateful to the anonymous referee for constructive comments.
K.H. is grateful to Noboru Kaneko, Atsuo~T. Okazaki,
James R. Murray, Yasuhiro Asano and 
Satoshi Tanda for their continuous encouragement. 
K.H. is also grateful to Atsuo~T. Okazaki, 
Jun Fukue and Shoji Kato
for helpful discussions.
Authors thank YITP in Kyoto University,
where this work was also extensively discussed
during the YITP-W-05-11
on September 20-21, 2005.
The simulations reported here were performed using the facility
of the Centre for Astrophysics \& Supercomputing at
Swinburne University of Technology, Australia and 
of YITP in Kyoto University.
This work has been supported 
in part by the Grants-in-Aid of the Ministry
of Education, Science, Culture, and Sport and Technology (MEXT;
14079205 K.H. \& S.M., 16340057 S.M.),
and by the Grant-in-Aid for the 21st Century COE
Scientific Research Programs on 
"Topological Science and Technology"
and
"Center for Diversity and Universality in Physics"
 from the MEXT.



\begin{thebibliography}{}
\bibitem[Abraham \& Carrara (1998)]{AC}
   Abraham,~Z., \& Carrara,~E.A. 1998, \apj, 496, 172
\bibitem[Armitage \& Natarajan (2002)]{armi1}
   Armitage,~P.J \& Natarajan,~P. 2002, \apj, 567, L9
\bibitem[Armitage \& Natarajan (2005)]{armi2}
   Armitage,~P.J \& Natarajan,~P. 2005, \apj, 634, 921
\bibitem[Artymowicz (1998)]{arty1}
   Artymowiz,~P. 1998, in Theory of Black Hole Accretion Discs, 
   ed. Abramowicz,~M.A., Bj\"ornsson,~G., \& Pringle,~J.E.
  (Cambridge: Cambridge University Press), 202
\bibitem[Artymowicz et al. (1991)]{arty2}
   Artymowicz,~P., Clarke,~C.L., Lubow,~H.S., \& Pringle,~J.E. 1991, \apj, 370, L35
\bibitem[Artymowicz \& Lubow (1994)]{arty3}
   Artymowicz,~P., \& Lubow,~H.S. 1994, \apj, 421, 651
\bibitem[Artymowicz \& Lubow (1996a)]{arty4}
   Artymowicz,~P., \& Lubow,~H. S. 1996a, \apj, 467, L77
\bibitem[Artymowicz \& Lubow (1996b)]{arty5}
   Artymowicz,~P., \& Lubow,~H.S. 1996b,
   In Disks and Outflows around Young Stars, ed. Beckwith,~S., Staude,~J., Quetz,~A., \& Natta,~A.  
   (Berlin, Heidelberg and NewYork: Springer-Verlag), 115
\bibitem[Bate \& Bonnell (1997)]{bate1}
   Bate,~M. R., \& Bonnel,~I.A. 1997, \mnras, 285, 33 
\bibitem[Bate et al.(1995)]{bate2}
   Bate,~M. R., Bonnel,~I. A., \& Price,~N.M. 1995, \mnras, 277, 362
\bibitem[Begelman et al.(1980)]{begel}
   Begelman,~M.C., Blandford,~R.D., \& Rees,~M.J. 1980, \nat, 287, 307 
\bibitem[Benz (1990)]{benz1}
   Benz W.\ 1990, in the Numerical Modeling of Nonlinear Steller Pulsations: Problems and Prospects,
   ed. Buchler,~R.J. 
   (Dordrecht: Kluwer Academic Publishers), 269 
\bibitem[Benz et al. (1990)]{benz2}
   Benz,~W., Bowers,~R.L., Cameron,~A.G.W., \& Press,~W.H. 1990, \apj, 348, 647 
\bibitem[Britzen et al.(2001)]{britzen}
   Britzen,~S., Roland,~J., Laskar,~J., Kokkotas,~K., Campbell,~R.M., \& Witzel,~A. 2001, \aap, 374, 784
\bibitem[Di Matteo et al.(2005)]{matteo}
   Di Matteo,~Tiziana, Springel,~V., \& Hernquist,~L. 2005, \nat, 433, 604
\bibitem[Ferrarese \& Merritt (2000)]{ferra}
   Ferrarese,~L., \& Merritt,~D. 2000, \apjl, 539, 9 
\bibitem[Gaskell (1996)]{gaskell}
   Gaskell,~C. Martin. 1996, \apjl, 464, 107
\bibitem[Gebhardt et al. (2000)]{geb}
   Gebhardt K \etal\ 2000, \apjl, 539, 13
\bibitem[G\"unther \& Kley (2002)]{GK1}
   G\"unther,~R. C., \& Kley,~W. 2002, \aap, 387, 550 
\bibitem[G\"unther \& Kley (2004)]{GK2}
   G\"unther,~R. C \& Kley,~W. 2004, \aap, 423, 559
\bibitem[Hayasaki \& Okazaki (2004)]{kimi1}
   Hayasaki,~K., \& Okazaki,~A.T.\ 2004, \mnras, 467, 77
\bibitem[Hayasaki \& Okazaki (2005)]{kimi2}
   Hayasaki,~K., \& Okazaki,~A.T.\ 2005, \mnras, 360, L15
\bibitem[Hayasaki \& Okazaki (2006)]{kimi3}
   Hayasaki,~K., \& Okazaki,~A.T. 2006, \mnras, 372, 1140
\bibitem[Ho et al. (2000)]{Ho}
   Ho,~C.Luis \etal\ 2000, \apj, 541, 120
\bibitem[Kataoka et al. (2003)]{kata}
   Kataoka, J., et al. 2003, \apj, 560, 659
\bibitem[Kato et al. (1998)]{kato}
   Kato,~S., Fukue,~J., \& Mineshige,~S. 1998, Black-Hole Accretion Disks
   (Kyoto: Kyoto University Press)
\bibitem[Katz (1997)]{katz}
   Katz,~J.I. 1997, \apj, 478, 527 
\bibitem[Kaufmann \& Haehnnelt (2000)]{KH}
   Kauffmann,~G.,\& Haehnelt,~M. 2000, \mnras, 311, 576 
\bibitem[Kitamura (1970)]{kitamura}
   Kitamura,~S. 1970, \apss, 7, 272 
\bibitem[Komossa (2003)]{komo1}
   Komossa,~S. 2003, Observational Evidence for Supermassive Black Hole Binaries,
   ed. Centrella,~J.M (AIP Conference Proceedings), 161
\bibitem[Komossa (2006)]{komo2}
   Komossa,~S. 2006, Memorie della Societa Astronomica Italliana, 77, 733
\bibitem[Kormendy \& Richstone (1995)]{KR}
   Kormendy,~J., \& Richstone,~D. 1995, \araa, 33, 581
\bibitem[Lehto \& Valtonen (1996)]{LV}
   Lehto,~H.J., \& Valtonen,~M.J. 1996, \apj, 460, 207
\bibitem[Lobanov \& Roland (2005)]{LR}
   Lobanov,~A.P., \& Roland,~J. 2005, \aap, 431, 831
\bibitem[Lubow \& Artymowicz (1996)]{lubow1}
   Lubow,~H. S., \& Artymowicz,~P. 1996, In Evolutionary 
   Processes in Binary Stars, ed. Ralph,~A.M., Wijers,~J., Melvyn,~B.D 
  (Dordrecht: Kluwer Academic Publishers), 53
\bibitem[Lubow\& Artymowicz (2000)]{lubow2}
   Lubow H. S., \& Artymowicz,~P. 2000, In Protostars
   and Planets IV, ed. Mannings,~V., Boss,~A.P., \& Russell,~S.S. 
   (Tucson: University of Arizona Press), 731
\bibitem[Lynden-Bell (1969)]{lynden}
   Lynden-Bell,~D. 1969, \nat, 223, 690
\bibitem[Miyoshi et al.(1995)]{miyoshi}
   Miyoshi,~M., Moran,~J., Herrnsteln,~J., 
   Greenhill,~L., Nakai,~N., Diamond,~P., \& Inoue,~M. 1995, \nat, 373, 127
\bibitem[Maness et al. (2004)]{maness}
   Maness,~H.L., Taylor,~G.B., Zavala,~R.T., Peck,~A.B., \& Pollack,~L.K. 2004, \apj, 602, 123
\bibitem[MacFadyen \& Milosavljevi\`c (2006)]{mac}
   MacFadyen,~A.I., \& Milosavljevi\'c,~M. 2006, astroph/0607467
\bibitem[Meglicki, Wickramanshige \& Bicknell (1993)]{MWB}
   Meglicki Z, Wickramasinghe D, Bicknell G\ 1993, \mnras, 264, 691
\bibitem[Merritt (2006)]{merritt1}
   Merritt David\ 2006, astroph/0605070
\bibitem[Merritt \& Ekers (2002)]{merritt2}
   Merritt,~D \& Ekers,~R.D. 2002, Science, 297, 1310
\bibitem[Milosavljevi\`c \& Merritt (2001)]{milos1}
   Milosavljevi\'c,~M., \& Merritt,~D. 2001, \apj, 563,34
\bibitem[Monaghan \& Gingold (1983)]{mona1}
   Monaghan J.J \& Gingold \ 1983, J.Comput.Phys, 52, 374
   ch.4, 160
\bibitem[Murray (1998)]{Murray}
   Murray,~J.R. 1998, \mnras, 297, 323
\bibitem[Negoro, Mineshige (2002)]{nego}
   Negoro, H.,\& Mineshige, S. 2002, \pasj, 54, L69
\bibitem[Ochi et al.(2005)]{ochi}
   Ochi Y, Sugimoto K \& Hanawa T\ 2005, \apj, 623, 922
\bibitem[Okazaki et al.(2002)]{oka}
   Okazaki,~A.T., Bate,~M.R., Ogilvie,~ G.I., \& Pringle J.E. 2002, \mnras, 337, 967
\bibitem[Papaloizou \& Pringle (1977)]{PP}
   Papaloizou,~J., \& Pringle,~J.E. 1977, \mnras, 181, 441 
\bibitem[Polnarev \& Rees (1994)]{PR}
   Polnarev, A.G., \& Rees,~M.J. 1994, \aap, 283, 301 
\bibitem[Quinlan \& Hernquist 1997]{QH}
   Quinlan,~G., \& Hernquist,~L. 1997, New Astron, 2, 533
\bibitem[Rauch \& Tremaine (1996)]{RT}
   Rauch,~K., \& Tremaine,~S. 1996, New Astron, 1, 149
\bibitem[Rees 1984]{rees}
   Rees M.J\ 1984, \araa, 22, 471
\bibitem[Rodriguez et al.(2006)]{rodriguez}
   Rodriguez C et al\ 2006, astroph/0604042
\bibitem[Roos 1981]{roos}
   Roos,~N. 1981, \aap, 104, 218
\bibitem[Roos et al. 1993]{roos2}
   Roos,~N., Kaastra,~J.S., \& Hummel,~C.A. 1993, \apj, 409, 130
\bibitem[Shakura \& Sunyaev (1973)]{SS}
   Shakura,~N.I., \& Sunyaev,~R.A. 1973, \aap, 24, 337
\bibitem[Silk \& Rees (1998)]{SR}
   Silk,~J., \& Rees,~M.J. 1998, \aap, 331, L1
\bibitem[Sillanp\"a\"a et al.(1988)]{sill}
   Sillanp\"a\"a,~A., Haarala,~S., Valtonen,~M., 
   Sundelius,~B., \& Byrdi,~G.G. 1988, \apj, 325, 628
\bibitem[Stothers \& Sillanp\"a\"a (1997)]{SSI}
   Stothers,~R.B., \& Sillanp\"a\"a,~A. 1997, \apjl, 475, L13
\bibitem[Sudou et al.(2003)]{sudou}
   Sudou,~H., Iguchi,~S., Muratai,~Y., \& Taniguchi,~T. 2003, Science, 300, 1263
\bibitem[Tanaka et al.(1995)]{tanaka}
   Tanaka,~Y., \etal\ 1995, \nat, 375, 659 
\bibitem[Valtaoja et al. (2000)]{valtao}
   Valtaoja,~E., \etal\ 2000, \apj, 531, 744
\bibitem[Valtonen et al.(2006)]{valtonen}
   Valtonen,~M.J., \etal\ 2006, \apjl, 643, L9
\bibitem[Yokosawa \& Inoue (1985)]{YI}
   Yokosawa,~M., \& Inoue,~M. 1985, \pasj, 37, 655
\bibitem[Yu (2002)]{yu}
   Yu,~Q. 2002, \mnras, 331, 935
\end{thebibliography}
\end{document}